\documentclass[12pt]{article}
\usepackage{longtable}
\usepackage{mathptm}
\usepackage{graphicx}
\usepackage{color}
\usepackage{pdfcolmk}	

\newcommand{\be}{\begin{equation}}
\newcommand{\ee}{\end{equation}}
\newcommand{\bea}{\begin{eqnarray}}
\newcommand{\eea}{\end{eqnarray}}
\newcommand{\degC}{{\,}^\circ C}
\newcommand{\bfi}{\begin{figure}[h!]}
\newcommand{\efi}{\end{figure}}

\def\preprint#1{ }			

\def\corr#1{#1}				

\def\url#1{\textcolor{blue}{\underline{#1}}}	

\preprint{ \large\normalsize}

\begin{document}


\thispagestyle{empty}

{\it Journal of Neuroscience Methods}, in press (2008).

\preprint{\vspace{12mm}}

\begin{center}

{\Large Characterizing synaptic conductance fluctuations in cortical
neurons \\ and their influence on spike generation \\ }

\

{\bf Zuzanna Piwkowska$^1$, Martin Pospischil$^1$, Romain Brette$^2$,
Julia Sliwa$^{1,3}$, \\ Michelle Rudolph-Lilith$^1$, Thierry Bal$^1$
and Alain Destexhe$^{1,*}$}

\preprint{\vspace{10mm}}

{\small 1: Unit\'e de Neurosciences Int\'egratives et
Computationnelles (UNIC), \\ CNRS, 91198 Gif-sur-Yvette, France

2: \'Equipe Odyss\'ee (ENS/INRIA/ENPC), \'Ecole Normale Sup\'erieure,
Paris, France

3: Present address: Centre de Neuroscience Cognitive (CNC), UMR 5229,
CNRS, 69675 Bron, France

*: corresponding author at address 1 above; \\ Tel: 33-1-6982-3435,
Fax: 33-1-6982-3427; destexhe@unic.cnrs-gif.fr}

\preprint{\vspace{20mm}}

\

{\bf Abbreviated title:} Characterizing synaptic conductance
fluctuations

\preprint{\vspace{10mm}}

\

Dynamic-clamp; Computational models;
Spike-triggered average; Inhibition

\

\end{center}


\preprint{\clearpage}

\begin{abstract}

Cortical neurons are subject to sustained and irregular synaptic
activity which causes important fluctuations of the membrane
potential (V$_m$). We review here different methods to characterize
this activity and its impact on spike generation. The simplified,
fluctuating point-conductance model of synaptic activity provides the
starting point of a variety of methods for the analysis of
intracellular V$_m$ recordings. In this model, the synaptic
excitatory and inhibitory conductances are described by
Gaussian-distributed stochastic variables, or ``colored conductance
noise''. The matching of experimentally recorded V$_m$ distributions
to an invertible theoretical expression derived from the model allows
the extraction of parameters characterizing the synaptic conductance
distributions. This analysis can be complemented by the matching of
experimental V$_m$ power spectral densities (PSDs) to a theoretical
template, even though the unexpected scaling properties of
experimental PSDs limit the precision of this latter approach. 
Building on this stochastic characterization of synaptic activity, we
also propose methods to qualitatively and quantitatively evaluate
spike-triggered averages of synaptic time-courses preceding spikes. 
This analysis points to an essential role for synaptic conductance
variance in determining spike times. The presented methods are
evaluated using controlled conductance injection in cortical neurons
{\it in vitro} with the dynamic-clamp technique. We review their
applications to the analysis of {\it in vivo} intracellular
recordings in cat association cortex, which suggest a predominant
role for inhibition in determining both sub- and supra-threshold
dynamics of cortical neurons embedded in active networks.  

\end{abstract}

\preprint{\vspace{10mm}}


\clearpage
\section{Introduction}

Cerebral cortical networks can generate states of intense and
irregular activity, which are characterized by low-amplitude
``desynchronized'' fast activity in the electroencephalogram (EEG), a
defining feature of the awake state. Intracellular measurements in
awake animals (Woody and Gruen, 1978; Matsumura et al., 1988; Baranyi
et al., 1993; Steriade et al., 2001; Timofeev et al., 2001; Rudolph
et al., 2007) have shown that cortical neurons are depolarized (about
-60 mV), have a low input resistance, their membrane potential
(V$_m$) fluctuates, and they fire irregularly and sustainedly. During
slow-wave sleep, or under several types of anesthetics (such as
urethane or ketamine-xylazine), the V$_m$ displays ``up-''
(depolarized) and ``down-'' (hyperpolarized) states, which are
paralleled with EEG slow waves (Metherate and Ashe, 1993; Steriade et
al., 1993; Steriade et al., 2001; Timofeev et al., 2001).  During the
up-state, the EEG is desynchronized and the V$_m$ of cortical neurons
is depolarized and highly fluctuating, similar to the sustained
activity found in awake animals (Destexhe et al., 2007). Up- and
down-states have also been found in ferret cortical slices using
high-potassium and low-calcium extracellular media (Sanchez-Vives and
McCormick, 2000) and in rat entorhinal slices as a function of the
metabolic state (Cunningham et al., 2006): these experiments indicate
that intracortical circuits are able to generate such states,
presumably through recurrent excitation and inhibition.  TTX block of
action potentials {\it in vivo} (Par\'e et al., 1998) and CNQX block
of excitatory synapses  {\it in vitro} (Cunningham et al., 2006)
abolish the depolarized, fluctuating states, which confirms their
synaptic origin.  

The recurrent activity of cortical networks has been investigated
using computational models at different levels and including various
degrees of biological detail. Large networks of formal (or simple
spiking) neurons allow the analytic derivation (or numerical
confirmation) of conditions for different classes of network
activity, such as oscillations or deterministic chaos (Van Vreeswijk
and Sompolinsky, 1996; Roxin et al., 2005; Barak and Tsodyks, 2007). 
Network models incorporating a realistic diversity of cell types and
details of cortical connectivity allow quantitative predictions to be
obtained through numerical simulations (Compte et al., 2003; Hill and
Tononi, 2005).  At the single neuron level, detailed models can
include the cell's morphology, a variety of intrinsic ion channels
and distributed synaptic inputs: studies of such models assess the
impact of massive input from the cortical network on dendritic
processing, spike train statistics, neuronal responsiveness and
integrative properties (Bernander et al., 1991; Destexhe and Par\'e,
1999; Rudolph and Destexhe 2003b,c; Destexhe et al., 2003b). However,
the parameterization of all those models requires a large amount of
information that cannot be obtained from any single experiment, which
complicates the comparison of simulation results with biological
data.  

A complementary approach consists in developing and studying models
that are relatively simple, but contain variables and parameters that
can be related to quantities directly measured in experiments. 
\corr{Various mathematical studies of the firing dynamics of neurons
with conductance-based synaptic inputs have been proposed (see for
example Burkitt et al., 2003; Moreno-Bote and Parga, 2005; Muller et
al., 2007). Other authors investigated the integration of multiple
current- or conductance-based inputs driven by Poisson spike trains,
in terms of V$_m$ fluctuations and output firing rate of simple point
model neurons (Kuhn et al., 2004). In the approach proposed by
Destexhe and colleagues (Destexhe et al., 2001), the effective impact
at the soma of thousands of single synapses, respectively excitatory
and inhibitory, activated by such Poisson spike trains, is
represented by only two stochastic conductance variables, $g_e(t)$
and $g_i(t)$, modeled as Ornstein-Uhlenbeck (Brownian-motion-like)
processes. This {\it fluctuating point-conductance model} describes
the evolution of the subthreshold V$_m$ of a single-compartment
neuron by a passive membrane equation with the two additional
stochastic conductance variables $g_e(t)$ and $g_i(t)$.} We present
here how this model can be used in close combination with
electrophysiological experiments to investigate the properties and
the impact of cortical recurrent activity at the single neuron level.

More specifically, we critically review, on the basis of new as well
as already published data, applications of the point-conductance
model of synaptic activity to the analysis of intracellular
recordings in cortical neurons: in each case, we first briefly
describe the method of analysis, then we show how it can be validated
experimentally by using controlled ``conductance injection'' in
biological neurons with dynamic-clamp, and finally we present the
application of the method to the analysis of real synaptic activity.
We especially focus on recently developed approaches for determining
which of the conductance configurations occurring in the fluctuating
synaptic activity trigger spikes.


\section{Methods}

\subsection{Computational methods}

Computational models were based on single-compartment neurons
described by the following membrane equation:
\be
 C \ \frac{dV}{dt}
  \ = \ - G_L \, (V-E_L)
    - g_e \, (V-E_e)
    - g_i \, (V-E_i)
    + \, I_{ext}			~ ,	\label{PCmodel}
\ee
where $C$ denotes the membrane capacitance, $I_{ext}$ a stimulation
current, $G_L$ the leak conductance and $E_L$ the leak reversal
potential.  $g_e(t)$ and $g_i(t)$ are stochastic excitatory
and inhibitory conductances, with respective reversal potentials 
$E_e$ and $E_i$.

These effective synaptic conductances were described by the following 
Ornstein-Uhlenbeck model (Destexhe et al., 2001):
\bea
 \frac{dg_e(t)}{dt} & = & 
 - \frac{1}{\tau_e} \, [ g_e(t) - g_{e0} ]
 + \sqrt{\frac{2 \sigma^2_e}{\tau_e}} \ \xi_e(t)     \label{flucte} \\
 \frac{dg_i(t)}{dt} & = & 
 - \frac{1}{\tau_i} \, [ g_i(t) - g_{i0} ]
 + \sqrt{\frac{2 \sigma^2_i}{\tau_i}} \ \xi_i(t) ~ , \label{flucti}
\eea
where $g_{e0}$ and $\sigma^2_e$ are, respectively, the mean value and
variance of the excitatory conductance, $\tau_e$ is the excitatory
time constant, and $\xi_e(t)$ is a Gaussian white noise source with
zero mean and unit standard deviation. The inhibitory conductance
$g_i(t)$ is described by an equivalent equation (Eq.~\ref{flucti})
with parameters $g_{i0}$, $\sigma^2_i$, $\tau_i$ and noise source
$\xi_i(t)$.  

In some simulations, the voltage-dependent conductances responsible
for action potentials, $g_{Na}$ and $g_{Kd}$ (with respective
reversals $E_{Na}$ and $E_K$), were included.  They were described by
Hodgkin-Huxley type models (with equations and parameters identical
as described in Destexhe et al., 2001, and references therein),
resulting in the following equation:
\be
 C \ \frac{dV}{dt}
  \ = \ - G_L \, (V-E_L)
    - g_{Na} \, (V-E_{Na})
    - g_{Kd} \, (V-E_K)
    - g_e \, (V-E_e)
    - g_i \, (V-E_i)
    + \, I_{ext}			~ .	\label{PCmodel2}
\ee

Simulations were performed on LINUX workstations using the NEURON
simulation environment (Hines and Carnevale, 1997).

\subsection{Biological preparation}

{\it In vitro} experiments were performed on 380-400 $\mu$m thick
coronal or sagittal slices from the lateral portions of
pentobarbital-anesthetized adult ferret (Marshall Europe, Lyon) and
guinea-pig (CPA, Olivet, France) occipital cortex, as described
previously (Rudolph et al., 2004; Pospischil et al., 2007). Slices
were maintained in an interface style recording chamber at
33-35$\degC$ in slice solution containing (in mM) 124 NaCl, 2.5 KCl,
1.2 MgSO$_4$, 1.25 NaHPO$_4$, 2 CaCl$_2$, 26 NaHCO$_3$, and 10
dextrose and aerated with 95\% O$_2$-5\% CO$_2$ to a final pH of 7.4.
In some experiments on ferret cortical slices, after approximately 1
hour, the solution was modified to contain 1 mM MgSO$_4$, 1 mM
CaCl$_2$ and 3.5 mM KCl (Sanchez-Vives and McCormick, 2000) in order
to obtain up-states.  Intracellular recordings following two hours of
recovery were performed in all cortical layers on
electrophysiologically identified regular spiking and intrinsically
bursting cells.  

All research procedures concerning the experimental animals and their
care adhered to the American Physiological Society's Guiding
Principles in the Care and Use of Animals, to the European Council
Directive 86/609/EEC and to European Treaties series no. 123, and was
also approved by the local ethics committee ``Ile-de-France Sud''
(certificate no. 05-003).

We also review data from intracellular recordings in cat association
cortex {\it in vivo}, which were described in detail elsewhere
(Rudolph et al., 2005; Steriade et al., 2001; Rudolph et al., 2007).

\subsection{Electrophysiology}

Sharp electrodes for intracellular recordings were made on a Sutter
Instruments P-87 micropipette puller from medium-walled glass (WPI,
1BF100) and beveled on a Sutter Instruments beveler (BV-10M). 
Micropipettes were filled with 1.2-2 M potassium acetate - 4 mM
potassium chloride and had resistances of 65-110 M$\Omega$ after
beveling.  An Axoclamp 2B amplifier (Axon Instruments) was used for
V$_m$ recording and current injection. A Digidata 1322A card (Axon
Instruments) was used for data acquisition at 20~kHz.

The dynamic-clamp technique (Robinson et al., 1993; Sharp et al.,
1993) was used to inject computer-generated conductances in real
neurons. Dynamic-clamp experiments were run as described previously
(Rudolph et al., 2004, Pospischil et al., 2007) using the hybrid
RT-NEURON environment (developed by G. Le Masson, INSERM
358, Universit\'e
Bordeaux~2), which is a modified version of NEURON (Hines and
Carnevale, 1997) running in real-time under the Windows 2000
operating system (Microsoft Corp.). In these experiments, the
injected current $I_{DynClamp}$ was determined from the fluctuating
conductances $g_e(t)$ and $g_i(t)$ modeled with Ornstein-Uhlenbeck
processes (Eqs.~\ref{flucte}--\ref{flucti}) as well as from the
difference between the recorded membrane voltage $V$ and the
respective reversal potentials:

\be
 I_{DynClamp} \  =  \ - g_e (V -E_e) \ - \ g_i (V -E_i) ~ .
\ee

The contamination of measured membrane voltage by electrode artefacts
was avoided either through the use of the discontinuous current-clamp
mode (in which current injection and voltage recording alternate at
frequencies of 2-3 kHz with our electrodes) or with Active Electrode
Compensation, a novel, high-resolution digital on-line compensation
technique we have recently developed (Brette et al., 2005, 2007;
Rudolph et al., 2005).

\subsection{Data analysis}

The PC-based software ELPHY (developed by G. Sadoc, CNRS
Gif-sur-Yvette, ANVAR and Biologic), Statview, Excel, custom-written
C-code and Neuron-code were used for analyses. All values are given
as average $\pm$ standard deviation.

\subsubsection{VmD analysis}

In dynamic-clamp experiments re-creating up-states with conductance
injection, two approaches were used. In 3 cells, conductance
estimates were computed for different, realistic values of leak
conductance and cell capacitance, and were then tested against the
real up-states in dynamic-clamp. In 2 cells, leak conductance and
cell capacitance were estimated from the response to short current
pulses, and those values were then used for conductance estimation:
the estimated conductance parameters, when used for dynamic-clamp
injection, proved to allow successful matching of the real up-state
in terms of V$_m$ distributions. In all cases, conductance estimation
was done during the recording using ELPHY.

VmD analysis of the {\it in vivo} data reviewed here is described in
Rudolph et al., 2005 and Rudolph et al., 2007.  

\subsubsection{Power Spectral Density (PSD) analysis}

Power spectra of V$_m$ activity were fit to an analytic template (see
Results) using a simplex fitting algorithm (Press et al., 1986). 
Different initial conditions (``first guesses'') were given to the
fitting procedure to ensure that there was no convergence to local
minima.  Some fits were realized by fitting both the amplitude of
excitatory and inhibitory components ($A_e$ and $A_i$; see Results),
as well as the time constants ($\tau_e$ and $\tau_i$).  In other
cases, it was not possible to fit 4 parameters from the experimental
PSD.  In such cases (typically from {\it in vivo} data), the fit was
performed with a single amplitude component ($A_e = A_i$).  In all
cases, the effective membrane time constant ($\tilde{\tau}_m$) was
fixed to the value estimated from the recordings.

\subsubsection{Spike-Triggered Average (STA) analysis}

For each conductance injection, spikes were detected using a
threshold at -30 mV. Inter-spike-intervals (ISIs) were computed and
for further analysis, a stable region in terms of ISI distribution
was used (as assessed by a non-significant Spearman correlation test
between ISI duration and time), including 673 $\pm$ 493 spikes. When
investigating the impact of frequency on the accuracy of the
estimates, the same number of spikes (52) was used for all analyzed
injections. Spikes following ISIs of at least 100~ms were then
selected. 50~ms-long pieces of V$_m$, $g_e$ and $g_i$ preceding each
selected spike were averaged to obtain the ``measured STAs''. 
\corr{Using the recorded conductance traces, we checked that
excluding spikes following ISIs shorter than 100~ms did not affect
the measured STAs in an important way (the difference in conductance
variation before the spike was of -0.14 $\pm$ 0.35~nS for excitation,
and of 0.5 $\pm$ 0.7~nS for inhibition).  However, in order to
compare with the STA extraction method based on the V$_m$ STA, it is
important to exclude short ISIs from the analysis to minimize
contamination of the V$_m$ STA by conductances related to preceding
spikes.}

For conductance STAs extraction based on the V$_m$ STA, parameters
were obtained in the following way: responses to depolarizing current
pulses were used to estimate the membrane capacitance from the time
constant of exponential fits to the decay of the V$_m$, and the leak
conductance was obtained by dividing the average voltage during
conductance injection, relative to rest, by the average injected
current. \corr{This results in an effective leak conductance possibly
comprising different types of voltage-dependent conductances. An
indication of the validity of this approach is provided by the
estimated conductance STAs: the important variation of this effective
leak conductance on the time scale of the length of the analysis
window (for example due to conductances underlying spike frequency
adaptation) would result in alterations of the stationary phase,
which we did not observe when excluding ISIs shorter than 100~ms. 
More details about the application of the STA analysis to {\it in
vivo} data can be found in Rudolph et al.\ (2007).} 

To quantify both the measured and the extracted conductance STAs, we
fitted the conductance time courses using the exponential template
\be
  g_e(t) = g^{STA}_{e0} \left[ 1 + k_e \ 
      \exp{(t-t_0) \over \tau^{STA}_{g_e}} \right] ~ ,
  \label{exp-template}
\ee
for excitation, and an equivalent equation for inhibition.  Here,
$t_0$ stands for the time of the spike, $k_e$ quantifies the maximal
increase/decrease of conductance prior to the spike ($\Delta g_e =
g_{e0} \ k_e$), with time constant $\tau^{STA}_{g_e}$, and 
$g^{STA}_{e0}$ is the average baseline conductance (see Results).


\section{Results}

The point-conductance model of recurrent cortical activity describes
the evolution of the subthreshold V$_m$ of a point neuron based on
two effective fluctuating conductances $g_e$ and $g_i$ (Destexhe et
al., 2001; see Methods for equations).  The basic assumption of this
model is that the synaptic conductances can be described as
Gaussian-distributed stochastic variables.  It was shown that the
well-known Ornstein-Uhlenbeck model of Brownian noise (Uhlenbeck and
Ornstein, 1930) approximates very well the total synaptic
conductances resulting from a large number of \corr{simulated}
conductance-based synaptic inputs (Destexhe et al., 2001; Destexhe
and Rudolph, 2004).  \corr{The fitting of a Gaussian model to the
total synaptic conductances seen at the soma, in simulations
performed using realistic cortical neuron morphologies and
distributed synaptic inputs,} indicates the following correspondences
between variables: the time constants ($\tau_e$, $\tau_i$) are
identical to the decay time constants of synaptic currents.  The
average conductance ($g_{e0}$, $g_{i0}$) is related to the overall
(integrated) conductance, which depends on the release frequency of
the corresponding Poisson inputs, the quantal conductance and the
decay time of synaptic currents.  The variance of the conductances
($\sigma^2_e$, $\sigma^2_i$) is related to the same parameters, as
well as to \corr{the amount of correlation between inputs of the same
type (Destexhe et al., 2001).  Correlations between presynaptic spike
trains seem necessary in order to account for the high amplitude of
V$_m$ fluctuations observed {\it in vivo} (e.g., Destexhe and Par\'e,
1999; L\'eger et al., 2005).}

\corr{It has to be noted that the Gaussian model of synaptic
conductances can only be considered as an approximation. As shown
theoretically using the shot noise formalism, in the presence of very
strong correlations, or a low mean presynaptic rate, this
approximation might fail (Richardson and Gerstner, 2005; Rudolph and
Destexhe, 2006; see also Kuhn et al., 2003). In this case, effective
synaptic conductances might be better described by a gamma
distribution.  However, the Gaussian model seems appropriate, even in
the presence of correlations, as long as the ratio of conductance
standard-deviation to conductance mean remains small (so that the
presence of negative conductances is negligible). Experimental
support for this assumption has been obtained using voltage-clamp
conductance measurements (Destexhe et al., 2003a).}

The theoretical and numerical analysis of the point-conductance model
has led to various useful derivations: an invertible expression for
the steady-state distribution of V$_m$ fluctuations (Rudolph and
Destexhe, 2003a, 2005), an expression for the power spectral density
of V$_m$ fluctuations (Destexhe and Rudolph, 2004), a geometrical
analysis of the configuration of conductances directly preceding
spikes, and a probabilistic method for calculating the most likely
conductance time course preceding spikes given an average V$_m$ time
course (Pospischil et al., 2007). We now proceed to examine in more
detail the applications of these computational results to the
analysis of intracellular recordings in cortical neurons.  

\subsection{The VmD method for extracting synaptic conductance
parameters}

\subsubsection{Outline of the VmD method}

The model described by Eqs.~\ref{PCmodel}--\ref{flucti} has been
thoroughly studied theoretically and numerically.  This model
describes the subthreshold V$_m$ fluctuations of a neuron subject to
fluctuating conductances $g_e$ and $g_i$.  Different analytic
approximations have been proposed to describe the steady-state
distribution of these V$_m$ fluctuations (Rudolph and Destexhe,
2003a, 2005; Richardson, 2004; Lindner and Longtin, 2006; for a
comparative study, see Rudolph and Destexhe, 2006).  One of these
expressions is invertible (Rudolph and Destexhe, 2003a, 2005), which
enables one to directly estimate the parameters ($g_{e0}$, $g_{e0}$,
$\sigma_e$, $\sigma_i$) from experimentally calculated V$_m$
distributions.  This constitutes the basis of the VmD method (Rudolph
et al., 2004), which we outline below.

The essential idea behind the VmD method is to fit an analytic
expression to the steady-state subthreshold V$_m$ distribution
obtained experimentally, and yield estimates of the parameters (mean,
variance) of the underlying synaptic conductances.  Among the
different analytic expressions outlined above, we consider the
following Gaussian approximation of the steady-state V$_m$ 
distribution: 
\be
\label{gaussf}
  \rho(V) \ \sim \ 
  \exp \left[ - \, \frac{(V - \bar{V})^2}{2 \, \sigma_V^2} \right] ~ ,
\ee
where $\bar{V}$ is the average V$_m$ and $\sigma_V$ its standard
deviation.  This expression provides an excellent approximation of
the V$_m$ distributions obtained from models and experiments (Rudolph
et al., 2004), because the V$_m$ distributions obtained
experimentally show little asymmetry (for up-states and activated
states; for specific examples, see Figs.~\ref{fig1} and \ref{fig2},
and Rudolph et al., 2004, 2005, 2007).  

One main advantage of this Gaussian approximation is that it can be
inverted, which leads to expressions of the synaptic noise parameters
as a function of the V$_m$ measurements, $\bar{V}$ and $\sigma_V$. 
By fixing the values of $\tau_e$ and $\tau_i$, which are related to
the decay time of synaptic currents and can be estimated from
voltage-clamp data and/or current-clamp by using power spectral
analysis (see Section~\ref{psdsec}), we remain with four parameters
to estimate: the means ($g_{e0}$, $g_{i0}$) and standard deviations
($\sigma_e$, $\sigma_i$) of excitatory and inhibitory synaptic
conductances.  To extract these four conductance parameters from the
membrane probability distribution, Eq.~\ref{gaussf} is, however,
insufficient because it is characterized by only two parameters
($\bar{V}$, $\sigma_V$).  To solve this problem, one possibility is
to consider two V$_m$ distributions obtained at two different
constant levels of injected current $I_{ext1}$ and $I_{ext2}$.  In
this case, the Gaussian approximation (Eq.~\ref{gaussf}) of the two
distributions gives two mean V$_m$ values, $\bar{V}_1$ and
$\bar{V}_2$, and two standard deviation values, $\sigma_{V1}$ and
$\sigma_{V2}$.  The resulting system of four equations relating V$_m$
parameters with conductance parameters can now be solved for four
unknowns: 

\bea
g_{ \{e,i\} 0}
 & = & \frac{  ( I_{ext1}-I_{ext2} ) \,
               \Big[  \sigma_{V2}^2 \left( E_{ \{i,e\} } - \bar{V}_1 
\right)^2
                    - \sigma_{V1}^2 \left( E_{ \{i,e\} } - \bar{V}_2 
\right)^2
               \Big]}
            {  \Big[  ( E_e - \bar{V}_1 ) ( E_i - \bar{V}_2 )
                    + ( E_e - \bar{V}_2 ) ( E_i - \bar{V}_1 )
               \Big]
               ( E_{ \{e,i\} } - E_{ \{i,e\} } ) 
               \left( \bar{V}_1 - \bar{V}_2 \right)^2 }         \label{gest} \\
 &   & 
     - \ \frac{
           ( I_{ext1}-I_{ext2} ) 
               ( E_{ \{i,e\} } - \bar{V}_2 )
             + \big[  I_{ext2}
                    - G_L ( E_{ \{i,e\} } - E_L ) \big]
               \left( \bar{V}_1^{\ } - \bar{V}_2^{\ } \right)
        }{
           ( E_{ \{e,i\} } - E_{ \{i,e\} } )
               \left( \bar{V}_1^{\ } - \bar{V}_2^{\ } \right)
        }
        \nonumber ~ ,
\eea

\bea
\sigma^2_{ \{e,i\} }
 & = & \frac{  2 C \ ( I_{ext1}-I_{ext2} ) \,
               \big[  \sigma_{V1}^2 \left( E_{ \{i,e\} } - \bar{V}_2 \right)^2
                    - \sigma_{V2}^2 \left( E_{ \{i,e\} } - \bar{V}_1 \right)^2
               \big] }
            {  \tilde{\tau}_{ \{e,i\} } \ 
               \Big[  ( E_e - \bar{V}_1 ) ( E_i - \bar{V}_2 )
                    + ( E_e - \bar{V}_2 ) ( E_i - \bar{V}_1 )
               \Big]
               ( E_{ \{e,i\} } - E_{ \{i,e\} } ) 
               \left( \bar{V}_1 - \bar{V}_2 \right)^2 }  ~ .    \label{sigest}
\eea
Here, $\tilde{\tau}_{ \{e,i\} }$ are effective time constants given
by (Rudolph and Destexhe, 2005): 
\be
  \tilde{\tau}_{\{e,i\}} 
 = \frac{2 \tau_{\{e,i\}} \tilde{\tau}_m}{\tau_{\{e,i\}} + \tilde{\tau}_m} 
 ~ ,  \label{tildetau}
\ee
where $\tilde{\tau}_m = C / (G_L + g_{e0} + g_{i0})$ is the effective
membrane time constant.

These relations enable us to estimate global characteristics of
network activity, such as mean excitatory ($g_{e0}$) and inhibitory
($g_{i0}$) synaptic conductances, as well as their respective
variances ($\sigma^2_e$, $\sigma^2_i$), from the sole knowledge of
the V$_m$ distributions obtained at two different levels of injected
current.  This VmD method was tested using computational models and
dynamic-clamp experiments (Rudolph et al., 2004) and was also used to
extract conductances from different experimental conditions {\it in
vivo} (Rudolph et al., 2005, 2007; Zou et al., 2005).

\subsubsection{Testing the VmD method with dynamic-clamp}

Taking advantage of the possibility, given by the dynamic-clamp
technique (see Methods), to mimic in a finely controlled way the
fluctuating conductances $g_e$ and $g_i$ in biological neurons, we
performed {\it in vitro} tests of the VmD method (Rudolph et al.,
2004; Piwkowska et al., 2004). In a first test (in 5
neurons), we computed V$_m$ distributions selectively during periods
of subthreshold activity collected within up-states recorded in
ferret cortical slices, we subsequently extracted conductance
parameters from Gaussian fits to these distributions, and finally we
used the estimated parameters to inject fluctuating conductances in
dynamic-clamp in the same cell, during down-states. Fig~\ref{fig1}C
shows a typical example of a real up-state and, shortly after, an
up-state re-created in dynamic-clamp. We confirmed that the V$_m$
distributions are very similar in the two cases (see Rudolph et al.,
2004 for more details). This test shows that the V$_m$ distributions observed
experimentally {\it in vitro} during recurrent cortical activity can
be accounted for by the proposed point-conductance model. We also
re-estimated known parameters of synaptic conductances (g$_{e0}$,
$g_{i0}$, $\sigma_e$, $\sigma_i$) injected in dynamic-clamp from the
resulting V$_m$ distributions: the match between actual and estimated
values is shown in Fig.~\ref{fig1}B. This second test indicates that
the passive approximation for the membrane behavior holds in the
studied case. In these tests, we did not consider the issue of the
estimation of $\tau_e$ and $\tau_i$ and assumed these values are
known.

\subsubsection{Analysis of intracellular recordings of cortical
neurons {\it in vivo}} \label{invivosec}

The VmD method was then applied to analyze intracellular recordings
in anesthetized (Rudolph et al., 2005), as well as naturally
sleeping and awake cats (Rudolph et al., 2007).  

In the first study, recordings were performed in cat association
cortex under ketamine-xylazine anesthesia, during the slow
oscillation typical of this anesthetic and resembling
slow-wave-sleep, as well as during prolonged periods of activity,
triggered by brain stem (PPT) stimulation, and with activity similar
to that of the aroused brain. The VmD method was used to extract
synaptic conductance parameters underlying the V$_m$ fluctuations of
the up-states of the slow oscillation, as well as those underlying
the continuous fluctuations following PPT stimulation. In both cases,
the average estimated inhibitory conductance $g_{i0}$ was markedly
higher than the average estimated excitatory conductance $g_{e0}$,
and similarly the estimated variance of inhibition $\sigma^2_i$ was
higher than the variance of excitation $\sigma^2_e$.  

The second study shows similar results across the natural wake-sleep
cycle of the cat (Fig.~\ref{fig2}): for a majority of cells,
especially during slow-wave-sleep up-states, inhibition dominated in
terms of both mean and variance. At the population level, the ratio
of inhibition to excitation was higher during slow-wave-sleep
up-states compared to the wake state. In 3 neurons that were recorded
across  several states, both average conductances together with their
variances decreased in the wake state compared to slow-wave-sleep
up-states. In addition, especially during the wake state, some cells
displayed comparable excitation and inhibition or even a dominant
excitation (2 out of 11 cells in the wake state). The study also
reports an important diversity in the absolute values of the
estimated conductance parameters.

An important concern in this type of studies is the estimation of the
leak parameters of the recorded neurons. The down-states of the slow
oscillation, when the local network is presumably silent, are too
short for properly estimating these parameters, since these brief
periods immediately following prolonged up-states are likely to
include after-hyperpolarizing currents (Sanchez-Vives et al., 2000)
that would bias the estimate of the leak.  In both cases, values
obtained from previous work (Par\'e et al, 1998; Destexhe and Par\'e,
1999) were used for the up-states: these studies evaluated that the
ratio of input resistance after TTX block of synaptic activity to the
total input resistance during up-states was about 4 to 6-fold. The
total input resistance was estimated in all cells studied from the
linear portion of the I-V curve obtained during up-states. During wake
or wake-like states, the mean total input resistance, as compared to
the mean total input resistance during up-states, was taken into
account to predict the ratio of input resistance with and without
synaptic activity. The underlying hypothesis for these assumptions is
that the ratio of leak conductance to synaptic conductance is similar
in all cells during comparable network states, implying that the leak
conductance and the total synaptic conductance covary in a strong way
across cells. In the natural wake-sleep cycle study, the dependence of
the conductance estimates on this important ratio was systematically
evaluated: this analysis showed that the estimated domination of
inhibition was qualitatively robust, and that this prediction failed
only when the leak conductance and the total synaptic conductance
were assumed to be approximately equal.  

On the other hand, the variance estimates do not depend on the leak
conductance, provided that the total conductance is known, as is the
case in the cited studies. They are, however, dependent on the
membrane capacitance $C$ (see Eqs.~\ref{sigest} and \ref{tildetau}). 
This value was assumed to be constant across the cells analyzed {\it
in vivo} (supposing a specific membrane capacitance of
1~$\mu$F/cm$^2$ and a membrane area of around 30000~$\mu$m$^2$). In
future studies, it could possibly be estimated using short current
pulse injection during the spontaneous activity, where the
capacitance can be extracted from the time constant of exponential
fits to the V$_m$ decay. Synaptic conductance variances are useful
parameters that can be related to spike initiation (see
Sections~\ref{stasec1} and \ref{stasec2}). The VmD method provided
the first published estimates for these parameters {\it in vivo} (for
a different approach, see Monier et al., 2008, this issue).

\subsection{Estimating time constants from V$_m$ power spectral density}
\label{psdsec}

\subsubsection{Outline of the method}

The point-conductance model given by Eqs.~\ref{PCmodel}--\ref{flucti}
was studied further, and recently we showed that the power spectral
density (PSD) of the V$_m$ fluctuations described by this model can
be well approximated by the following expression (Destexhe and
Rudolph, 2004):
\be
  S_V(\omega) \ = \ { 4 \over G_T^2} \ 
              { 1 \over 1 + \omega^2 \ \tilde{\tau}_m^2 } \
              \left[ {\sigma_e^2 \tau_e \ (E_e-\bar{V})^2 \over 1 + \omega^2 \ \tau_e^2} \ + \ 
                     {\sigma_i^2 \tau_i \ (E_i-\bar{V})^2 \over 1 + \omega^2 \ \tau_i^2} 
              \right]            ~ ,		\label{VmPSD}
\ee
where $\omega$ = $2 \pi f$, $f$ is the frequency, $G_T = G_L + g_{e0}
+ g_{i0}$ is the total membrane conductance, $\tilde{\tau}_m = C /
G_T$ is the effective time constant, and $\bar{V} = ( G_L E_L +
g_{e0} E_e + g_{i0} E_i ) / G_T$ is the average membrane potential. 
The ``effective leak'' approximation used to derive this equation
consisted in incorporating the average synaptic conductances into the
total leak conductance, and then considering that fluctuations around
the obtained mean voltage are subjected to a constant driving force (Destexhe
and Rudolph, 2004).

As mentioned above, the synaptic time constant parameters, $\tau_e$
and $\tau_i$, need to be estimated in order to extract precise values
of the conductance variances with the VmD method. As those two
parameters appear in the theoretical expression of the V$_m$ PSD, we
explored the possibility of evaluating them from the analysis of the
PSD of experimentally recorded V$_m$ fluctuations. To this end,
the following simplified expression can be fitted:
\be
  S_V(\omega) \ = \ 
              { 1 \over 1 + \omega^2 \ \tilde{\tau}_m^2 } \
              \left[ {A_e \ \tau_e \over 1 + \omega^2 \ \tau_e^2} \ + \ 
                     {A_i \ \tau_i \over 1 + \omega^2 \ \tau_i^2} 
              \right]            ~ ,		\label{VmPSD2}
\ee
where $A_e$ and $A_i$ are amplitude parameters.  This five parameter
template is used to provide estimates of the parameters $\tau_e$ and
$\tau_i$ (supposing that $\tilde{\tau}_m$ has been measured).  A
further simplification consists in assuming that $A_e = A_i$, which
was used for fitting {\it in vivo} data (see Methods).

These analytic expressions were tested by comparing the prediction to
numerical simulations of a single-compartment model subject to
fluctuating synaptic conductances (Eqs.~\ref{PCmodel}--\ref{flucti}).
The matching between the analytic expression and the PSD obtained
numerically was nearly perfect, as shown in Fig.~\ref{fig3}A and as
reported in detail previously (Destexhe and Rudolph, 2004).  

\subsubsection{Testing synaptic time constants estimates with
dynamic-clamp}

We applied the procedure described above to the PSD of V$_m$
fluctuations obtained by controlled, dynamic-clamp fluctuating
conductance injection in cortical neurons {\it in vitro} (using a
new, high resolution electrode compensation technique, Brette et al.,
2005, 2007; Rudolph et al., 2005).  In this case, the scaling of the
PSD conforms to the prediction (Fig.~\ref{fig3}B): the theoretical
template (Eq.~\ref{VmPSD2}) can provide a very good fit of the
experimentally obtained PSD, up to around 400~Hz, where recording
noise becomes important.   The template used was according to
Eq.~\ref{VmPSD} or Eq.~\ref{VmPSD2}, both of which provided equally
good fits (not shown).  This shows that the analytic expression for
the PSD is consistent not only with models, but also with conductance
injection in real neurons {\it in vitro}.

\subsubsection{PSD Analysis of V$_m$ fluctuations {\it in vitro} and
{\it in vivo}}

We have also attempted to apply the same procedure to V$_m$
fluctuations resulting from real synaptic activity, during up-states
recorded {\it in vitro} (Fig.~\ref{fig3}C) and during sustained
network activity {\it in vivo} (Fig.~\ref{fig3}D).  In this case,
however, it is apparent that the experimental PSDs cannot be fitted
with the theoretical template as nicely as for dynamic-clamp data
(Fig.~\ref{fig3}B).  The PSD presents a frequency scaling region at
high frequencies, and scales as $1/f^\alpha$ with a different
exponent $\alpha$ as predicted by the theory (see
Figs.~\ref{fig3}C-D).  The analytic expression (Eq.~\ref{VmPSD})
predicts that the PSD should scale as 1/f$^4$ at high frequencies,
but the experiments show that the exponent $\alpha$ is obviously
lower than that value (see Discussion for possible reasons for such a
difference).  This difference of course compromises the accuracy of
the method to estimate $\tau_e$ and $\tau_i$ in situations of real
synaptic bombardment.  Nevertheless, including the values of $\tau_e$
= 3~ms and $\tau_i$ = 10~ms provided acceptable fits to the
low-frequency ($<$100~Hz) part of the spectrum (Fig.~\ref{fig3}C-D,
red curves).  However, in this case, small variations (around
20-30\%) around these values of $\tau_e$ and $\tau_i$ yielded equally
good fits (not shown; see also Rudolph et al., 2005).  Thus, the
method cannot be used to precisely estimate those parameters, but can
nevertheless be used to broadly estimate them with an error of the
order of 30~\%.

\subsection{Estimating spike-triggering conductance configurations}
\label{stasec1}

\subsubsection{A preliminary investigation}

The conductance measurements outlined above show that there is a
diversity of combinations of $g_e$ and $g_i$ that underlies the
genesis of subthreshold activity in different preparations. We used a
computational model based on the point-conductance model, but
including a spiking mechanism (see Eq.~\ref{PCmodel2}), to reproduce
those measurements and try to infer what different properties such
states may have in terms of spike selectivity.  Indeed, an infinite
number of combinations of $g_e$ and $g_i$ can give similar V$_m$
activity.  Figure~\ref{fig4}A illustrates two extreme examples out of
this continuum: first a state where both excitatory and inhibitory
conductances are of comparable magnitude (Fig.~\ref{fig4}A, left;
``Equal conductances''). In this state, both conductances are lower
than the resting conductance of the cell and the V$_m$ is fluctuating
around -60~mV. Second, similar V$_m$ fluctuations can be obtained
when both conductances are of larger magnitude, but in this case,
inhibition has to be augmented several-fold to maintain the V$_m$
around -60~mV (Fig.~\ref{fig4}A, right; ``Inhibition-dominated''). 
Such conductance values are more typical of what is usually measured
{\it in vivo} \corr{(Rudolph et al., 2005, 2007; Monier et al., 2008,
this issue)}. Both conductances are larger than the resting
conductance, a situation which can be described as a
``high-conductance state''.

To determine how these two states differ in their spike selectivity,
we evaluated the spike-triggering conductances by averaging the
conductance traces collected in 50~ms windows preceding spikes. This
average pattern of conductance variations leading to spikes is shown
in Fig.~\ref{fig4}B. For equal-conductance states, there is an
increase of total conductance preceding spikes (purple curve in
Fig.~\ref{fig4}B, left), as can be expected from the fact that
excitation increases ($g_e$ curve in Fig.~\ref{fig4}B, left). In
contrast, for inhibition-dominated states, the total conductance
decreases prior to the spike (purple curve in Fig.~\ref{fig4}B,
right), and this decrease necessarily comes from a similar decrease
of inhibitory conductance, which is, in this case, stronger than the
increase of excitatory conductance ($g_i$ curve in Fig.~\ref{fig4}B,
right). Thus, in such states the spike seems primarily caused by a
drop of inhibition.

This pattern was seen not only in the average, but also at the level
of single spikes. Using a vector representation to display the
conductance variation preceding spikes (each vector links the
conductance state in a window of 30-40 ms before the spike with that
in the 10 ms preceding the spike) shows that the majority of spikes
follow the average pattern (Fig.~\ref{fig4}C).   The same features
were also present when the integrate-and-fire model was used (not
shown), and thus do not seem to depend on the spike generating
mechanisms.

These patterns of conductance variations preceding spikes were also
investigated in real neurons by using dynamic-clamp experiments to
inject fluctuating conductances {\it in vitro}. In this case, performing the same analysis
 as above revealed similar features: spike-triggered averages (STAs) 
of the injected conductances displayed either increase or decrease in total conductance,
 depending on the conductance parameters used (Fig.~\ref{fig5}A), and the vector representations were
 also similar (Fig.~\ref{fig5}B). It suggested that these features are independent of the spike generating
 mechanism but rather are caused by subthreshold V$_m$ dynamics.

\subsubsection{A geometrical interpretation based on the
point-conductance model}

The configuration of synaptic conductances just before spikes can be
explained qualitatively by considering that the total current must be
positive at spike time, i.e., $g_e(E_e-V_t) + g_i(E_i-V_t) +
G_L(E_L-V_t)>0$, where $V_t$ is the spike threshold (using an
integrate-and-fire approximation).

This inequality defines a half-plane in which $(g_e, g_i)$ must lie
at spike time. Fig.~\ref{fig6}A shows graphically how this inequality
affects the synaptic conductances. The variable $(g_e,g_i)$ is
normally distributed, so that isoprobability curves are ellipses in
the plane (plotted in red). In that plane, the line
$\{g_e+g_i=g_{e0}+g_{i0}\}$ going through the center of the ellipses
defines the points for which the total conductance equals the mean
conductance, and the line $\left\{g_e(E_e-V_t) + g_i(E_i-V_t) +
G_L(E_L-V_t)=0\right\}$ defines the border of the half-plane in which
conductances lie at spike time.  In the equal conductances regime
(Fig.~\ref{fig6}A, left), synaptic conductances are small and have
similar variances, so that isoprobability curves are circular; the
intersection of the half-plane with those circles is mostly above the
mean total conductance line, so that the total conductance is higher
than average at spike time.

In the inhibition-dominated regime (Fig.~\ref{fig6}A, right), synaptic
conductances are large and the variance of $g_i$ is larger than the
variance of $g_e$, so that isoprobability curves are vertically
elongated ellipses; the intersection of the half-plane with those
ellipses is essentially below the mean total conductance line, so
that the total conductance is lower than average at spike time.

More precisely, when isoprobability curves are circular (equal
variances), then the expected total conductance is unchanged at spike
time when the lines $\left\{g_e(E_e-V_t) + g_i(E_i-V_t) +
G_L(E_L-V_t)=0\right\}$ and $\{g_e+g_i=g_{e0}+g_{i0}\}$ are
orthogonal, i.e., when $E_e-V_t+E_i-V_t=0$. Spikes are associated
with increases in conductance when the first line has a higher slope,
i.e., when $E_e-V_t>V_t-E_i$ (which is typically the case).  

When isoprobability curves are not circular, we can look at the graph
in the space $(\frac{g_e}{\sigma_e},\frac{g_i}{\sigma_i})$ where
isoprobability curves are circular. Then the orthogonality condition
between the lines
$$\left\{\frac{g_e}{\sigma_e}\sigma_e(E_e-V_t)+\frac{g_i}{\sigma_i}
\sigma_i(E_i-V_t)+G_L(E_L-V_t)=0\right\}$$ and
$$\left\{\frac{g_e}{\sigma_e}\sigma_e+\frac{g_i}{\sigma_i}\sigma_i
=g_{e0}+g_{i0}\right\}$$ 
reads
$$\sigma_e^2(E_e-V_t)+\sigma_i^2(E_i-V_t)=0 ~ .$$  It follows that
spikes are associated with increases in total conductance when the
following condition is met:
$$
\frac{\sigma_e}{\sigma_i}>\sqrt{\frac{V_t-E_i}{E_e-V_t}} ~ .
$$

One can also recover this result by calculating the expectation of
the conductance change conditionally to the fact that the current at
spike threshold is positive (implicitly, we are neglecting the
correlation time constants of the synaptic conductances). Using
typical values ($V_t=-55$ mV, $E_e=0$ mV, $E_i=-75$ mV), we conclude
that spikes are associated with increases in total conductance when
$\sigma_e>0.6\sigma_i$. This inequality is indeed satisfied in the
equal conductances regime and not in the inhibition-dominated regime
investigated above.  

\subsubsection{Testing the geometrical prediction with dynamic-clamp}

The geometrical reasoning predicts that the sign of the total
synaptic conductance change triggering spikes depends only on the
ratio of synaptic variances, and not on the average conductances. We
have systematically tested this prediction using dynamic-clamp
injection of fluctuating conductances {\it in vitro}. In 8 regular
spiking cortical neurons, we scanned different parameter regimes in a
total of 36 fluctuating conductance injections. Fig.~\ref{fig6}B
shows two examples from the same cell: both correspond to an average
``high conductance'' regime, dominated by inhibition, but in one case
it is the variance of excitation, in the other case the variance of
inhibition, that is higher. We can see that the total conductance
before the spike increases in the first case, but decreases in the
second. Fig.~\ref{fig6}C (left) shows the average total conductance
change preceding spikes as a function of $\sigma_e$/$\sigma_i$, for
all the 36 injections: the vertical dashed line represents the
predicted value of $\sigma_e$/$\sigma_i$=0.6, which indeed separates
all the ``conductance drop'' configurations from the ``conductance
increase'' configurations. Even though the prediction is based on a
simple integrate-and-fire extension of the point-conductance model,
we can see that the ratio of synaptic variances can predict the sign
of the total conductance change triggering spikes in biological
cortical neurons subjected to fluctuating excitatory and inhibitory
conductances.  

In addition (Fig.~\ref{fig6}C, right), the dynamic-clamp data shows
that the average amplitude of change ($\Delta g_e = g_{e0} \ k_e$,
see Methods) of each synaptic conductance preceding a spike is
related, in a linear way, to the standard deviation of this
conductance. For a fixed value of standard deviation, there was no
significant influence of the average conductance (not shown). This
observation is consistent with the idea that in all the cases studied
here, the firing of the cell was driven by fluctuations in the V$_m$,
rather than by a high mean V$_m$ value (not shown).

Taken together, these theoretical and experimental analyses indicate
that the average total conductance drop preceding spikes, as seen in
the ``high conductance'' case we initially considered
(Fig.~\ref{fig4}), is not a direct consequence of the ``high
conductance'' state of the membrane, but is in fact related to the
high inhibitory variance, which is indeed to be expected especially
when the mean inhibitory conductance is also high (as confirmed by
the studies presented in the first part of this article).

\subsection{Estimating spike-triggered averages of synaptic
conductances from the V$_m$}  \label{stasec2}

In order to extend the analysis of spike-triggering conductance
configurations to real recurrent cortical activity observed {\it in
vivo}, we recently developed a procedure to extract the
spike-triggered averages (STAs) of conductances from recordings of
the V$_m$ (Fig.~\ref{fig7}; Pospischil et al., 2007).  The STA of the
V$_m$ is calculated first, and the method searches for the ``most
likely'' spike-related conductance time courses ($g_e(t)$, $g_i(t)$)
that are compatible with the observed voltage STA. The procedure is
based on a discretization of the time axis in
Eqs.~\ref{PCmodel}--\ref{flucti}, which, rearranged, lead to the
following relations:

\bea
    \label{eq:gi}
    g_i^k &=& -\frac{C}{V^k-E_i} \left\{ \frac{V^k-E_L}{\tau_L} + 
    \frac{g_e^k (V^k - E_e)}{C} + \frac{V^{k+1} - V^k}{\Delta t} -
    \frac{I_{ext}}{C} \right\}, \\
    \label{eq:xi_e}
    \xi_e^k &=& \frac{1}{\sigma_e} \sqrt{\frac{\tau_e}{2 \Delta t}}
    \left( g_e^{k+1} - g_e^k \Big( 1 - \frac{\Delta t}{\tau_e} \Big) - 
    \frac{\Delta t}{\tau_e} g_{e 0} \right), \\  
    \label{eq:xi_i}
    \xi_i^k &=& \frac{1}{\sigma_i} \sqrt{\frac{\tau_i}{2 \Delta t}} 
    \left( g_i^{k+1} - g_i^k \Big( 1 - \frac{\Delta t}{\tau_i} \Big) - 
    \frac{\Delta t}{\tau_i} g_{i 0} \right) ~ ,
\eea
where $\tau_L = C/G_L$ is the resting membrane time constant.  Note
that $\xi_e(t)$ and $\xi_i(t)$ have become Gaussian--distributed
random numbers $\xi_e^k$ and $\xi_i^k$.  There is a continuum of
combinations $\{ g_e^{k+1}, g_i^{k+1}\}$ that can advance the
membrane potential from $V^{k+1}$ to $V^{k+2}$, each pair occurring
with a probability
\begin{eqnarray}
    p^k &:=& p(g_e^{k+1}, g_i^{k+1} | g_e^k, g_i^k) = \frac{1}{2 \pi} e^{-\frac{1}{2} (\xi_e^{k 2} + \xi_i^{k 2})} = \frac{1}{2 \pi} e^{-\frac{1}{4 \Delta t} X^k}, \\
    X^k &=& \frac{\tau_e}{\sigma_e^2} \left( g_e^{k+1} - g_e^k \Big( 1 - \frac{\Delta t}{\tau_e} \Big) - \frac{\Delta t}{\tau_e} g_{e 0} \right)^2 \\ 
    && + \frac{\tau_i}{\sigma_i^2} \left( g_i^{k+1} - g_i^k \Big( 1 - \frac{\Delta t}{\tau_i} \Big) - \frac{\Delta t}{\tau_i} g_{i 0} \right)^2. \nonumber
\end{eqnarray}
Because of Eq.~\ref{eq:gi},  $g_e^k$ and $g_i^k$ are not independent
and $p^k$ is, thus, a unidimensional distribution only. Given initial
conductances $\{g_e^0, g_i^0\}$, one can write down the probability
$p$ for certain series of conductances $\{g_e^j, g_i^j\}_{j = 0,
\ldots, n}$ to occur that reproduce a given voltage trace $\{ V^l
\}_{l = 1, \ldots, n+1}$:
\begin{equation}
    p = \prod_{k = 0}^{n-1} p^k.
\end{equation}
Due to the symmetry of the distribution $p$, the average paths of the
conductances coincide with the most likely ones. It is thus
sufficient to determine the conductance series with extremal
likelihood by solving the n-dimensional system of linear equations
\begin{equation}
    \label{eq:sys}
    \left\{ \frac{\partial X}{\partial g_e^k} = 0\right\}_{k = 1, \ldots, n},
\end{equation}
where $X = \sum_{k = 0}^{n-1} X^k$, for the vector $\{ g_e^k \}$. 
This is equivalent to solving $\{\frac{\partial p}{\partial g_e^k} =
0\}_{k = 1, \ldots, n}$ and involves the numerical inversion of an $n
\times n$-matrix, which can be done using standard numeric methods
(Press et al., 1986).  The series $\{ g_i^k \}$ is subsequently
obtained from Eq.~\ref{eq:gi}.  Details of this procedure as well as
an evaluation of its performance can be found in Pospischil et al.,
2007.

The method requires first an estimation of the parameters describing
the distribution of each of the conductances, which can be obtained
by the VmD method. The leak parameters of the cell, or alternatively
the effective parameters during V$_m$ fluctuations (effective
conductance and effective time constant), also have to be estimated
prior to this analysis.  The method was successfully tested using
computational models: conductance STAs extracted from the V$_m$ of an
integrate-and-fire point-conductance model are nearly identical to
the numerically obtained conductance STAs (Fig.~\ref{fig7}A, left;
Pospischil et al., 2007).  

During response to sensory stimuli, there can be a substantial degree
of correlation between excitatory and inhibitory synaptic input
(Monier et al., 2003; Wehr and Zador, 2003; Wilent and Contreras,
2005).  Since this situation has not been addressed in Pospischil et
al., 2007, we would like to sketch a possible extension of the
method. To this end, we reformulate the discretized versions of
Eqs.~\ref{flucte}, \ref{flucti} in the following way:
\bea
    \label{eq:ge_corr}
    \frac{g_e^{k+1} - g_e^{k}}{\Delta t} &=& -\frac{g_e^{k} - g_{e0}}{\tau_e} + \sigma_e \sqrt{\frac{2 \Delta t}{\tau_e (1+c)}} (\xi_1^k + \sqrt{c} \, \xi_2^k), \\
    \label{eq:gi_corr}
    \frac{g_i^{k+1} - g_i^{k}}{\Delta t} &=& -\frac{g_i^{k} - g_{i0}}{\tau_i} + \sigma_i \sqrt{\frac{2 \Delta t}{\tau_i (1+c)}} (\xi_2^{k-d} + \sqrt{c} \, \xi_1^{k-d}).
\eea

Here, instead of having one ``private'' white noise source feeding
each conductance channel, now the same two noise sources $\xi_1$ and
$\xi_2$ contribute to both inhibition and excitation.  The amount of
correlation is tuned by the parameter $c$.  Also, since there is
evidence that the peak of the $g_e$--$g_i$--crosscorrelation is not
always centered at 0 during stimulus-evoked responses (``delayed
inhibition''; see Wehr and Zador, 2003; Wilent and Contreras, 2005),
we allow a non--zero delay $d$: for a positive parameter $d$, the
inhibitory channel receives the input that the excitatory channel
received $d$ time steps before. Eqs.~\ref{eq:ge_corr} and
\ref{eq:gi_corr} can be solved for $\xi_1^k$ and $\xi_2^k$, thus
replacing Eqs.~\ref{eq:xi_e} and \ref{eq:xi_i}. It is then possible
to proceed as in the uncorrelated case, where now, due to the delay,
the matrix describing Eq.~\ref{eq:sys} has additional subdiagonal
entries.

However, the application of this extended method requires the
estimation of the usual leak parameters, of conductance distribution
parameters -- for which the VmD method cannot be directly used in its
current form since it is based on uncorrelated noise sources -- as
well as knowledge of the parameters $c$ and $d$. At present, we can
only speculate on how $c$ and $d$ could be evaluated in experiments:
extracellularly recorded spike trains could perhaps be used to this
end, provided that simultaneously recorded single units could be
classified as excitatory or inhibitory. Alternatively, different
plausible $c$ and $d$ values could be scanned to examine how they
could potentially influence the conductance STAs extracted from a
given V$_m$ STA.

\subsubsection{Testing STA estimation with dynamic-clamp}

The dynamic-clamp data presented above was used to evaluate the
accuracy of the conductance STA estimation method (for the
uncorrelated case only): indeed, the conductance STAs estimated from
the V$_m$ STAs could be compared to conductance STAs obtained
directly by averaging conductance traces, since in dynamic-clamp the
injected conductances are perfectly controlled by the
experimentalist. For the 36 injections analyzed, a good match was
observed between the two (see one example in Fig.~\ref{fig7}A, right
panel). Since we intended to evaluate the STA method specifically,
we assumed that conductance distribution parameters were known. In
addition, we estimated the cell's leak parameters (see Methods). To
quantify the comparison on a population basis, we did the following
analyses: exponential functions (see Methods) were fitted to each
conductance STA (estimated and directly measured) starting at 1~ms
before the spike and decaying  to baseline backwards in time.  We
then compared the asymptotic values (i.e., the average baseline
conductances) and the time constants of these fits, as well as the
amplitude of conductance change from the start of the fit to the
asymptote (i.e., the amplitude of conductance change preceding the
spike).  In all cases but one, the exponential functions provided
excellent fits to the conductance STAs. It was necessary to exclude
the 1~ms time window preceding the spike to avoid severe
contamination of the analyses by intrinsic conductances. Excluding a
broader time window did not improve the analyses for these neurons.

The average baseline conductances always matched very well (error of
-0.1 $\pm$ 0.25~nS, or -0.8 $\pm$ 2.6 \%, for excitation; 0.8 $\pm$
1.5~nS, or 0.6 $\pm$ 4.5\%,  for inhibition; not shown). More
importantly, the estimates of the average conductance patterns
leading to spikes were also in good correspondence with the measured
patterns, in terms of both the amplitude of conductance change
(Fig.~\ref{fig7}C, top; error of -1.2 $\pm$ 3~nS, or -26 $\pm$ 28.8
\%, for excitatory amplitude change, and -2.0 $\pm$ 2.5~nS, or -10.7
$\pm$ 47\%, for inhibitory amplitude change) and the time constant
(Fig.~\ref{fig7}C, bottom; error of 0.39 $\pm$ 0.48 ms, or 11.2 $\pm$
21.1\% for excitatory time constant, and 0.36 $\pm$ 1.71, or 2.6
$\pm$ 18.8\%, for inhibitory time constant).  For excitation, the
error on the estimate of the amplitude is correlated with the error
on the estimate of the time constant (not shown): this suggests that,
in most cases, a slightly too fast rise of the excitatory conductance
results in a slightly too high amplitude of conductance change. The
lack of correlation between the two error measures in the case of the
inhibitory conductance points to a more complex origin for the
observed errors. Moreover, the errors on the amplitude of the two
conductance changes are positively correlated (Fig.~\ref{fig7}D,
top). This dependency actually ensures that the error on the
estimated total conductance change (excitation-inhibition) remains
small (-0.8 $\pm$ 2.4~nS; Fig.~\ref{fig7}D, bottom).  

Finally, we have investigated the dependency of the estimate errors
on a diversity of variables, and found a correlation of amplitude
errors with the average V$_m$ during the fluctuating conductance
injection (not shown): this dependency points to a possible
contamination by intrinsic conductances activated differentially at
different average V$_m$ levels. However, we found no dependency of
the error on average firing rate (for the rates up to around 30 Hz
studied here), suggesting no important contamination by
spike-dependent conductances like the ones underlying the after-hyperpolarization (when
care is taken to compute STAs using spikes preceded by at least 100~ms
of silence, see Methods).

We have also verified that conductance STA estimates can be relied on
to investigate what factors determine the average conductance
variations preceding spikes. Fig.~\ref{fig8} shows that the analyses
performed previously on dynamic-clamp data (i.e., on STAs obtained
directly by averaging the conductance traces, Fig.~\ref{fig6}B-C),
can also be successfully performed using the conductance STAs
estimated from the corresponding V$_m$ STAs: Fig.~\ref{fig8}A shows
the estimated STAs in the two different ``high conductance'' states,
dominated by either excitatory or inhibitory variance (compare to
Fig.~\ref{fig6}B).  Fig.~\ref{fig8}B shows, for the population data,
the (significant) correlations between total conductance change and
$\sigma_e$/$\sigma_i$, as well as between the change of each of the
conductances and the corresponding standard deviation.  Note the
similarity between Fig.~\ref{fig6}B-C and Fig.~\ref{fig8}A-B, even
though the correlations at the population level are more noisy when
the estimated STAs are used.

\subsubsection{STA Analysis of intracellular recordings of cortical
neurons {\it in vivo}}

The conductance STA estimation method was used to determine
conductance variations preceding spikes during V$_m$ fluctuations
{\it in vivo} (Rudolph et al., 2007). Starting from V$_m$ recordings
of spontaneous spiking activity in awake or naturally sleeping cats,
we computed the spike-triggered average of the V$_m$
(Fig.~\ref{fig9}).  Using values of $g_{e0}$, $g_{i0}$, $\sigma_e$,
$\sigma_i$ estimated using the VmD method (see above), we computed
the most likely conductance traces yielding the observed V$_m$
averages.  Most of these analyses (7 out of 10 cells for awake, 6 out
of 6 for slow-wave-sleep, 2 out of 2 for REM) revealed conductance
dynamics consistent with states dominated by inhibitory variance:
there was a drop of the total conductance preceding spikes, due to a
strong decrease of the inhibitory conductance (Fig.~\ref{fig9},
right).  However, a few cases, in the wake state (3 out of 10 cells),
displayed the opposite configuration with the total synaptic
conductance increasing before the spike (Fig.~\ref{fig9}, left).

We also checked how the geometrical prediction relating the sign of
total conductance change preceding spikes and the ratio
$\sigma_e$/$\sigma_i$ performed for this data (Fig.~\ref{fig9}B). We
have seen that the critical value of $\sigma_e$/$\sigma_i$ for which
the total conductance change shifts from positive to negative depends
on the spike threshold. This parameter was quite variable in the
recorded cells, and so a critical $\sigma_e$/$\sigma_i$ value was
calculated for each cell. Fig.~\ref{fig9}B shows the lowest and
highest critical values obtained (dashed lines), and also displays in
white the cells which do not conform to the prediction based on their
critical value.  This is the case for only 4 out of 18 cells, for
three of which the total conductance change is close to zero.  

The extraction of conductance STAs depends on the accuracy of the
synaptic conductance parameters estimated with the VmD method, which
means that assumptions made about the leak conductance and the cell
capacitance will influence the results (see Section~\ref{invivosec}).
We have shown that the ratio $\sigma_e$/$\sigma_i$ should determine
whether, on average, the total conductance increases or decreases
prior to the spike.  This ratio is independent of the leak
conductance, but it depends on the capacitance $C$. However, it
appears in both the numerator and the denominator of the ratio. 
Fig.~\ref{fig9}C shows the dependency of $\sigma_e$/$\sigma_i$ on this
parameter for different values of total input resistance and two sets
of realistic values for the V$_m$ distribution parameters. This
analysis indicates that a reasonable error on $C$ produces a limited
error on the ratio $\sigma_e$/$\sigma_i$, and suggests that
conclusions drawn from the {\it in vivo} data about the respective
contributions of excitation and inhibition in triggering spikes are
valid.


\section{Discussion}

\subsection{\corr{Synaptic conductance analysis methods based on the
fluctuating point-con\-ductance model}}

We presented how the simple point-conductance model of cortical
synaptic activity can provide a basis for the analysis of
experimental data, essentially through the matching of expressions
derived from the model to intracellular V$_m$ recordings of cortical
neurons. This approach has been used for extracting different
parameters from the recurrent cortical activity {\it in vivo}: the
averages and variances of excitatory and inhibitory conductances,
their decay time \corr{constants} and the optimal conductance
waveform underlying spike selectivity.  These analyses were possible
because the point conductance model represents in a compact and
mathematically tractable way the activity resulting from several
thousand synapses.  

\corr{The VmD analysis} provides a characterization of synaptic
activity in simple terms (average conductance, level of
fluctuations).  Such parameters can readily be incorporated in
computational models to yield the V$_m$ and conductance state
corresponding to {\it in vivo} activity with just a few variables. 
This approach has been used for example in network simulations to
obtain realistic conductance states in neurons even with small
networks (Haeusler and Maass, 2007).  It is also directly usable in
dynamic-clamp experiments to investigate the impact of synaptic
background activity on signal processing by single cortical or
thalamic neurons (Fellous et al., 2003;  Shu et al., 2003; Wolfart et
al., 2005; Desai and Walcott, 2006).

Beyond the matching of experimental V$_m$ distributions to a
theoretical expression (VmD method), we have also attempted to match
the PSDs of V$_m$ fluctuations.  This approach provides some
validation for assumptions made about synaptic time constants on the
basis of published studies (Destexhe and Par\'e, 1999; Destexhe et al., 2001).
  However, the fact that the point-conductance model
does not account for the scaling properties of experimental PSDs
(Fig.~\ref{fig3}C-D) limits the accuracy of the method and yields
only broad estimates of the synaptic time constants (approximately
30\% error).  A parallel study (Bedard and Destexhe, 2007) has shown
that the frequency scaling observed experimentally cannot be
accounted for by standard cable theory, and modifications of cable
equations are required to match those values.  We hope to obtain a
more accurate fitting template, which would allow more precise PSD
analyses in the future.  

\corr{Finally, in order to be able to study average spike-triggered
patterns of conductances {\it in vivo}, we have recently developed a
probabilistic method for extracting the STAs of conductances from
STAs of the V$_m$ (Pospischil et al., 2007). This method relies on
the estimation of a number of parameters inherent to the neuron (like
leak conductance and capacitance), as well as on synaptic conductance
parameters estimated, for example, with the VmD method. We provided
new data from dynamic-clamp experiments {\it in vitro}, demonstrating
that given the synaptic conductance parameters, a good match between
the extracted conductance STAs and the actual, known conductance STAs
can be obtained.}

\subsection{\corr{Comparison to other synaptic conductance analysis
methods}}

Other studies proposed synaptic conductance estimates {\it in vivo}
derived from I-V curves (in current-clamp) or V-I curves (in
voltage-clamp) obtained at different points in time following a
stimulus (Borg-Graham et al.,1998; Anderson et al., 2000; Monier et
al., 2003; Wehr and Zador, 2003; Wilent and Contreras, 2005) or the
onset of an up-state (Haider et al., 2006). An exhaustive comparison
of these studies is beyond the scope of the present article (see
Monier et al., 2008, this issue), but a few points can be stressed. 
The fact that the VmD method applies to current-clamp data obtained
at a few levels of constant injected current circumvents technical
problems with voltage-clamp due to high series resistance of patch
electrodes {\it in vivo} (but see Borg-Graham et al., 1998; Monier et
al., 2003; Wehr and Zador, 2005) or the need for discontinuous
voltage-clamp with sharp electrodes (Haider et al., 2006). The fact
that it relies on a strong assumption about the stochasticity of
synaptic inputs and the independence of excitation and inhibition
allows the analysis of spontaneous cortical activity with no ``zero"
time point. However, this assumption also makes it unsuited, in its
present form, for the analysis of stimulus-evoked activity with
important temporal structure.

The VmD method suffers from one common limitation
with other approaches for synaptic conductance estimation:
the need to separate synaptic currents from leak currents. The fact
that it is used for the analysis of on-going, spontaneous
activity in cortical networks poses, however, an additional complication:
indeed, when synaptic inputs evoked
by sensory stimulation are analyzed, it is with reference to the pre-stimulus
activity, which includes the leak current and any other baseline
currents, including on-going synaptic activity (Borg-Graham et
al.,1998; Anderson et al., 2000; Monier et al., 2003;
Wehr and Zador, 2003; Wilent and Contreras, 2005). When we attempt to
analyze the spontaneous activity itself, there is no straightforward reference that can be
used. As mentioned above, the short down-states do not seem a good
candidate since they include after-hyperpolarizing currents
consecutive to the up-states (Sanchez-Vives et al., 2000), although a down-state-referenced
analysis (as in Haider et al., 2006) could perhaps be compared to the analysis performed in the
reviewed studies.  However, in the wake state, the continuous
on-going activity does not even present down-states. The only precise
approach to evaluate the leak conductance for each studied cell is to
block all synaptic activity with TTX, but that also means recording
only one or two cells per animal for this protocol, which is an
extremely constraining experimental situation.  The compromise chosen
in the reviewed {\it in vivo} studies consisted in using published
average values obtained in previous TTX experiments in a similar
preparation (Par\'e et al., 1998) and checking the robustness of the
estimates to the assumed leak conductance parameter. The
estimates should be re-evaluated in the light of any future
experimental data providing information about the leak parameters of
cortical cells {\it in vivo} in the absence of synaptic activity. 
Optimally, such future studies would include the simple, classical
protocol required for a subsequent VmD analysis: several-second-long
current-clamp recordings of the V$_m$ at different levels of steady
injected current.

\corr{As to our probabilistic method for extracting the STAs of
conductances from STAs of the V$_m$ (Pospischil et al., 2007),} we
are not aware of any other method currently allowing this analysis
for spontaneous activity: in voltage-clamp, no spikes are recorded
and so, obviously, conductances leading to spikes cannot be extracted
directly.  Voltage-clamp can only be used for extracting plausible
conductance STAs when the precise times of spikes are known and
reproducible (for example, at a given delay after a sensory
stimulation), so that the estimated, stimulus-locked conductance
dynamics can be reasonably expected to lead to spikes in the
current-clamp configuration (Monier et al., 2003; Wehr and Zador,
2003).

\subsection{\corr{Dependency of the STA extraction method on the VmD
method}}

The result of the STA analysis is, however, dependent on the
estimates of synaptic averages and variances, so that we may ask to
what extent this result depends on the accuracy of the VmD method. In
general, the average baseline conductances will reflect the estimates
of average synaptic conductances (since both analyses are constrained
by the same total input resistance measure and the same leak
conductance assumption), unless the total input resistance changes
markedly between the current levels used for the VmD analysis and the
zero-current level at which STAs are extracted, due to activation of
intrinsic conductances: in this case, since the conductances ($G_L$,
$g_e(t)$, $g_i(t)$) have to be compatible with the voltage $V_m(t)$,
the conductance STA baselines can be considerably shifted away from
the mean conductance values tens of ms before the spike. The same
effect is seen in {\it in vitro} dynamic-clamp experiments when a
wrong leak conductance value is used (not shown). This distortion
could be used as an indication for important activation of intrinsic
conductances and suggest that the result should be discarded.

\corr{The variances of the synaptic conductances are correlated to
the amplitudes of average conductance change preceding a spike (as
shown in Fig.~\ref{fig6}).  The estimation of synaptic conductance
variances with the VmD method is independent of the assumption made
about the leak conductance, which excludes this source of potential
error. It is dependent on the membrane capacitance $C$, which should
be evaluated on a cell-by-cell basis whenever possible in future
studies. However, we have also shown that the ratio of synaptic
conductance variances is only weakly dependent on the precise value
of $C$: this ratio determines the sign of total conductance change
preceding a spike, so that the estimates of this sign from {\it in
vivo} data seem robust (see below).} 

\subsection{\corr{Dynamic-clamp as a tool to evaluate conductance
analysis methods}}

We illustrated how and to what extent the validity of different
approaches for conductance analysis can be tested using
dynamic-clamp. This electrophysiological technique is an attractive
tool to evaluate methods of conductance analysis, since it allows to
mimic the activation of known conductances in a biological neuron:
the results of an analysis method based on V$_m$ recordings can be
directly compared to measures of the actual conductances controlled
by the experimentalist. In all the dynamic-clamp applications
presented here, we have used the same description for the synaptic
conductances - the Ornstein-Uhlenbeck stochastic model - as in the
theoretical analyses. This means that we could compare how the
analysis methods perform if the stochastic conductances are inserted
at the soma of a real cortical neuron, with a complex structure and a
variety of intrinsic channels, instead of a passive single
compartment.  As we have seen, potential dendritic effects solicited
only during distributed synaptic stimulation, like, possibly, the
unexpected scaling of the V$_m$ PSDs during real synaptic activity
(Bedard and Destexhe, 2007), cannot be addressed with this somatic
injection technique. They could perhaps be investigated in the future
using dendritic patch-clamp.  

The comparison was performed most extensively for the conductance STA
estimation method: it indicates that if the window of analysis is
chosen properly (by excluding a window of about 1~ms before the
spike, and also excluding inter-spike-intervals shorter than around
100~ms), the estimations perform well, and that the estimation
errors, correlated with the average V$_m$, are presumably linked to
V$_m$-dependent intrinsic channels.  However, we did not
systematically compare how the methods perform if the synaptic
conductances deviate from the Ornstein-Uhlenbeck model: such an
approach could constitute another application of the dynamic-clamp
tool to the evaluation of conductance analysis methods.  We have also
not attempted yet to evaluate in dynamic-clamp the extended STA
analysis method sketched above, which incorporates a known
correlation between excitation and inhibition
(Eqs.~\ref{eq:ge_corr}--\ref{eq:gi_corr}).

\subsection{\corr{Patterns of excitation and inhibition triggering
spikes depend on the variances of the synaptic inputs}}

What are the patterns of excitation and inhibition triggering spikes
under different conditions of network activity~?  A preliminary
study, \corr{using both models and dynamic-clamp conductance
injection {\it in vitro}}, pointed to the fact that these patterns
depend on the statistics of synaptic conductances, and that spikes
could be preceded, on average, either by increases in total synaptic
conductance, indicating a predominant role for excitation, or by
decreases in total synaptic conductance, indicating a predominant
role for inhibition. Other authors (Hasenstaub et al., 2005) have
recently suggested that drop of inhibition can play an important role
in determining spike timing in cortical neurons, based on
dynamic-clamp injection of specific synaptic conductances with
parameters matched to their {\it in vivo} recordings. Here we
explored this issue further and showed, first by a theoretical
reasoning, and second by scanning different parameter regimes using
dynamic-clamp conductance injection, that the sign of the total
conductance change before a spike does not directly depend on the
average synaptic conductances, but is solely determined by the ratio
of synaptic conductance variances. The variance of synaptic
conductances in the Ornstein-Uhlenbeck model is related to the degree
of correlation between Poisson input trains of each type in a
detailed biophysical model (Destexhe et al., 2001): the level of
synchrony among inhibitory neurons could thus be determining for
spike timing whenever it significantly exceeds the level of synchrony
among excitatory neurons. \corr{The rule relating the ratio of
synaptic conductance variances with the sign of the average
conductance change preceding a spike appears as a generalization of
more specific results obtained by Tateno and Robinson in cortical
neurons (Tateno and Robinson, 2006), as well as by the group of
Jaeger in the cerebellum (Gauck and Jaeger, 2000, 2003; Suter and
Jaeger, 2004): these authors also used dynamic-clamp injection of
synaptic conductances and the conductance STAs they obtained seem
consistent with our rule.}

\corr{This result stresses the importance of evaluating the variances
of synaptic conductances, in addition to their averages, when
analyzing V$_m$ fluctuations recorded {\it in vivo}. To our
knowledge, the two reviewed studies using the VmD method (Rudolph et
al., 2005, 2007) are the only ones (together with Monier et al.,
2008, this issue) explicitely providing estimates of synaptic
conductance variances. The application of the probabilistic method
also allowed for the first time the extraction of synaptic
conductance STAs from {\it in vivo} recordings of spontaneous V$_m$
fluctuations in awake and naturally sleeping cats (Rudolph et al.,
2007), and led to the observation of both types of firing regimes
described above -- average total conductance increase and average
total conductance drop -- with a majority of cases displaying the
inhibition-dominated, conductance-drop pattern.}


\subsection*{Acknowledgments}

We thank Igor Timofeev for giving us permission to use some of his
data.  Research supported by CNRS, ANR, ACI, HFSP and the European
Community (FACETS grant FP6 15879).  Z.P. gratefully acknowledges the
support of the FRM.

\preprint{\clearpage}

\section*{References}

\begin{description}

\small

\item Anderson JS, Carandini M, Ferster D. Orientation tuning of
input conductance, excitation, and inhibition in cat primary visual
cortex. J Neurophysiol 2000;84:909-26.

\item Barak O, Tsodyks M. Persistent activity in neural networks with
dynamic synapses. PLoS Comput Biol 2007;3:e35.

\item Baranyi A, Szente MB, Woody CD.  Electrophysiological
characterization of different types of neurons recorded in vivo in
the motor cortex of the cat. II. Membrane parameters, action
potentials, current-induced voltage responses and electrotonic
structures.  J  Neurophysiol 1993;69:1865-79.

\item Bedard C, Destexhe A. A modified cable formalism for modeling
neuronal membranes at high frequencies.  Biophys. J. 2007; in press. 
Preprint available at \url{http://arxiv.org/abs/0705.3759}.

\item Bernander O, Douglas RJ, Martin KA, Koch C.  Synaptic
background activity influences spatiotemporal integration in single
pyramidal cells. Proc Natl Acad Sci USA 1991;88:11569-73.  

\item Borg-Graham LJ, Monier C, Fr\'egnac Y.  Visual input evokes
transient and strong shunting inhibition in visual cortical neurons. 
Nature 1998;393:369-73.

\item Brette R, Rudolph M, Piwkowska Z, Bal T, Destexhe A. How to
emulate double-electrode recordings with a single electrode? A new
method of active electrode compensation. Soc Neurosci Abstracts
2005;688.2

\item \corr{Brette R, Piwkowska Z, Rudolph M, Bal T, Destexhe, A. A
nonparametric electrode model for intracellular recording.
Neurocomputing 2007;70:1597-601.}

\item \corr{Burkitt AN, Meffin H, Grayden DB. Study of neuronal gain
in a conductance-based leaky integrate-and-fire neuron model with
balanced excitatory and inhibitory synaptic input.  Biol. Cybern. 
2003;89:119-25.}

\item Compte A, Sanchez-Vives MV, McCormick DA, Wang XJ. Cellular and
network mechanisms of slow oscillatory activity ($<$1~Hz) and wave
propagations in a cortical network model. J Neurophysiol
2003;89:2707-25.

\item Cunningham MO, Pervouchine DD, Racca C, Kopell NJ, Davies CH,
Jones RS, Traub RD, Whittington MA. Neuronal metabolism governs
cortical network response state. Proc Natl Acad Sci USA
2006;103:5597-601.

\item Desai NS, Walcott EC. Synaptic bombardment modulates muscarinic
effects in forelimb motor cortex. J Neurosci 2006;26:2215-26.

\item Destexhe A and Par\'e D. Impact of network activity on the
integrative properties of neocortical pyramidal neurons in vivo.  J
Neurophysiol 1999;81:1531-47.

\item \corr{Destexhe A, Badoual M, Piwkowska Z, Bal T, Hasenstaub A,
Shu Y, McCormick DA, Pelletier J, Par\'e D, Rudolph M.  In vivo, in
vitro and computational evidence for balanced or inhibition-dominated
network states, and their respective impact on the firing mode of
neocortical neurons.  Soc. Neurosci. Abstracts 2003a;29:921.14}

\item Destexhe A, Rudolph M. Extracting information from the power
spectrum of synaptic noise. J Comput Neurosci 2004;17:327-45.

\item Destexhe A, Hughes SW, Rudolph M, Crunelli V.  Are
corticothalamic ``up'' states fragments of wakefulness? Trends
Neurosci 2007; doi:10.1016/j.tins.2007.04.006

\item Destexhe A, Rudolph M, Fellous J-M, Sejnowski TJ. Fluctuating
synaptic conductances recreate in vivo-like activity in neocortical
neurons. Neuroscience 2001;107:13-24.

\item Destexhe A, Rudolph M and Par\'e D.  The high-conductance state
of neocortical neurons in vivo.  Nature Reviews Neurosci
2003b;4:739-51.

\item Fellous JM, Rudolph M, Destexhe A, Sejnowski TJ. Synaptic
background noise controls the input/output characteristics of single
cells in an in vitro model of in vivo activity.  Neuroscience
2003;122:811-29.

\item \corr{Gauck V, Jaeger D. The control of rate and timing of
spikes in the deep cerebellar nuclei by inhibition. J. Neurosci. 
2000;20:3006-16.}

\item \corr{Gauck V, Jaeger D. The contribution of NMDA and AMPA
conductances to the control of spiking in neurons of the deep
cerebellar nuclei. J. Neurosci. 2003;23:8109-18.}

\item Haider B, Duque A, Hasenstaub AR, McCormick DA. Neocortical
network activity in vivo is generated through a dynamic balance of
excitation and inhibition. J Neurosci 2006;26:4535-45.

\item Hasenstaub A, Shu Y, Haider B, Kraushaar U, Duque A, McCormick
DA. Inhibitory postsynaptic potentials carry synchronized frequency
information in active cortical networks. Neuron 2005;47:423-35.

\item Haeusler S and Maass W. A statistical analysis of
information-processing properties of lamina-specific cortical
microcircuit models. Cereb Cortex 2007;17:149-62.

\item Hill S, Tononi G. Modeling sleep and wakefulness in the
thalamocortical system. J Neurophysiol 2005;93:1671-98.

\item Hines ML, Carnevale NT. The NEURON simulation environment,
Neural Computation 1997;9:1179-209.

\item \corr{Kuhn A, Aertsen A, Rotter S.  Higher-order statistics of
input ensembles and the response of simple model neurons.  Neural
Comput. 2003;15:67-101.}

\item \corr{Kuhn A, Aertsen A, Rotter S. Neuronal integration of
synaptic input in the fluctuation-driven regime.  J. Neurosci. 
2004;24:2345-56.}

\item \corr{Leger J-F, Stern EA, Aertsen A, Heck D. Synaptic
integration in rat frontal cortex shaped by network activity. J. 
Neurophysiol. 2005;93:281-93.}

\item Lindner B, Longtin A. Comment on ``Characterization of
subthreshold voltage fluctuations in neuronal membranes'', by M. 
Rudolph and A. Destexhe. Neural Comput 2006;18:1896-931.

\item Matsumura M, Cope T, Fetz EE.  Sustained excitatory synaptic
input to motor cortex neurons in awake animals revealed by
intracellular recording of membrane potentials.  Exp Brain Res
1988;70:463-69.

\item Metherate R, Ashe JH. Ionic flux contributions to neocortical
slow waves and nucleus basalis-mediated activation: whole-cell
recordings in vivo. J Neurosci 1993;13:5312-23.

\item \corr{Monier C, Fournier J, Fr\'egnac Y. In vitro and in vivo
measures of evoked excitatory and inhibitory conductance dynamics in
sensory cortices.  J. Neurosci. Meth. 2008; in press.}

\item Monier C, Chavane F, Baudot P, Graham LJ, Fr\'egnac Y.
Orientation and direction selectivity of synaptic inputs in visual
cortical neurons: a diversity of combinations produces spike tuning.
Neuron 2003;37:663-80.

\item \corr{Moreno-Bote R, Parga N. Membrane potential and response
properties of populations of cortical neurons in the high conductance
state.  Phys Rev Lett. 2005;94:088103.}

\item \corr{Muller E, Buesing L, Schemmel J, Meier K. 
Spike-frequency adapting neural ensembles: beyond mean adaptation and
renewal theories.  Neural Computation 2007;19:2958-3010.}

\item Par\'e D, Shink E, Gaudreau H, Destexhe A, Lang EJ. Impact of
spontaneous synaptic activity on the resting properties of cat
neocortical neurons in vivo.  J Neurophysiol 1998;79:1450-60.

\item Piwkowska Z, Rudolph M, Badoual M, Destexhe A, Bal T.
Re-creating active states in vitro with a dynamic-clamp protocol.
Neurocomputing 2005;65-66:55-60.

\item Pospischil M, Piwkowska Z, Rudolph M, Bal T, Destexhe A.
Calculating event-triggered average synaptic conductances from the
membrane potential. J Neurophysiol 2007;97:2544-52.

\item Press WH, Flannery BP, Teukolsky SA, Vetterling WT. Numerical
Recipes. The Art of Scientific Computing.  Cambridge, MA: Cambridge
University Press; 1986.

\item Richardson MJ. Effects of synaptic conductance on the voltage
distribution and firing rate of spiking neurons. Phys Rev E Stat
Nonlin Soft Matter Phys  2004;69:051918.

\item \corr{Richardson MJ, Gerstner W.  Synaptic shot noise and
conductance fluctuations affect the membrane voltage with equal
significance.  Neural Comput. 2005;17:923-47.}

\item Robinson HP, Kawai N. Injection of digitally synthesized
synaptic conductance transients to measure the integrative properties
of neurons. J Neurosci Methods 1993;49:157-65.

\item Roxin A, Brunel N, Hansel D. Role of delays in shaping
spatiotemporal dynamics of neuronal activity in large networks. Phys
Rev Lett 2005;94:238103.

\item Rudolph M, Destexhe A.  Characterization of subthreshold
voltage fluctuations in neuronal membranes. Neural Computation
2003a;15:2577-618.

\item Rudolph M, Destexhe A. The discharge variability of neocortical
neurons during high-conductance states.  Neuroscience
2003b;119:855-73.

\item Rudolph M, Destexhe A. A fast-conducting, stochastic
integrative mode for neocortical neurons in vivo. J.  Neurosci
2003c;23:2466-76.

\item Rudolph M, Destexhe A. An extended analytic expression for the
membrane potential distribution of conductance-based synaptic noise.
Neural Computation 2005;17:2301-15.

\item Rudolph M, Destexhe A. On the use of analytic expressions for
the voltage distribution to analyze intracellular recordings.  Neural
Comput. 2006; 18: 2917-22.

\item \corr{Rudolph M, Destexhe A.  A multichannel shot noise
approach to describe synaptic background activity in neurons. Eur.
Physical J.  2006; B 52: 125-32.}

\item Rudolph M, Piwkowska Z, Badoual M, Bal T, Destexhe A. A method
to estimate synaptic conductances from membrane potential
fluctuations. J Neurophysiol 2004;91:2884-96.

\item Rudolph M, Piwkowska Z, Brette R, Destexhe A, and Bal T.
Precise dynamic-clamp injection of stochastic conductances using
active electrode compensation.  Soc Neurosci Abstracts 2005;687.13.

\item Rudolph M, Pelletier J-G, Par\'e D, Destexhe A. 
Characterization of synaptic conductances and integrative properties
during electrically-induced EEG-activated states in neocortical
neurons in vivo. J Neurophysiol 2005;94:2805-21.

\item \corr{Rudolph M, Pospischil M, Timofeev I, Destexhe A. 
Inhibition determines membrane potential dynamics and controls action
potential generation in awake and sleeping cat cortex.  J. Neurosci. 
2007;27:5280-90.}

\item Sanchez-Vives MV, McCormick DA. Cellular and network mechanisms
of rhythmic recurrent activity in neocortex. Nat Neurosci
2000;3:1027-34.

\item Sanchez-Vives MV, Nowak LG, McCormick DA.  Cellular mechanisms
of long-lasting adaptation in visual cortical neurons in vitro.  J
Neurosci 2000;20:4286-99.  

\item Sharp AA, O'Neil MB, Abbott LF, Marder E. Dynamic clamp:
computer-generated conductances in real neurons. J Neurophysiol
1993;69:992-5.

\item Shu Y, Hasenstaub A, Badoual M, Bal T, McCormick DA. Barrages
of synaptic activity control the gain and sensitivity of cortical
neurons. J Neurosci 2003;23:10388-10401.

\item Steriade M, Nunez A, Amzica F. A novel slow ($<$ 1~Hz)
oscillation of neocortical neurons in vivo: depolarizing and
hyperpolarizing components. J Neurosci 1993;13:3252-65.

\item Steriade M, Timofeev I, Grenier F.  Natural waking and sleep
states: a view from inside neocortical neurons. J Neurophysiol
2001;85:1969-85.

\item \corr{Suter KJ, Jaeger D. Reliable control of spike rate and
spike timing by rapid input transients in cerebellar stellate cells. 
Neuroscience 2004;124:305-17.}

\item \corr{Tateno T, Robinson HP. Rate coding and spike-time
variability in cortical neurons with two types of threshold dynamics.
J. Neurophysiol. 2006;95:2650-63.}

\item Timofeev I, Grenier F, Steriade M.  Disfacilitation and active
inhibition in the neocortex during the natural sleep-wake cycle: an
intracellular study. Proc Natl Acad Sci USA 2001;98:1924-29.

\item Uhlenbeck GE and Ornstein LS, On the theory of the Brownian
motion.  Phys Rev 1930;36:823-41.

\item van Vreeswijk C, Sompolinsky H. Chaos in neuronal networks with
balanced excitatory and inhibitory activity. Science 1996;274:1724-26.

\item Wehr M, Zador AM. Balanced inhibition underlies tuning and
sharpens spike timing in auditory cortex.  Nature 2003;426:442-6.

\item Wehr M, Zador AM. Synaptic mechanisms of forward suppression in
rat auditory cortex.  Neuron 2005;47:437-45.

\item Wilent W, Contreras D. Dynamics of excitation and inhibition
underlying stimulus selectivity in rat somatosensory cortex.  Nature
Neurosci 2005;8:1364-70.

\item Wolfart J, Debay D, Le Masson G, Destexhe A, Bal T. Synaptic
background activity controls spike transfer from thalamus to cortex. 
Nat Neurosci 2005;8:1760-67.

\item Woody CD, Gruen E.  Characterization of electrophysiological
properties of intracellularly recorded neurons in the neocortex of
awake cats: a comparison of the response to injected current in spike
overshoot and undershoot neurons.  Brain Res 1978;158:343-57.

\item Zou Q, Rudolph M, Roy N, Sanchez-Vives M, Contreras D, Destexhe
A. Reconstructing synaptic background activity from conductance
measurements in vivo.  Neurocomputing 2005;65:673-78.

\end{description}


\clearpage

\bfi 
\section*{Figures}
\centerline{\includegraphics[width=14cm]{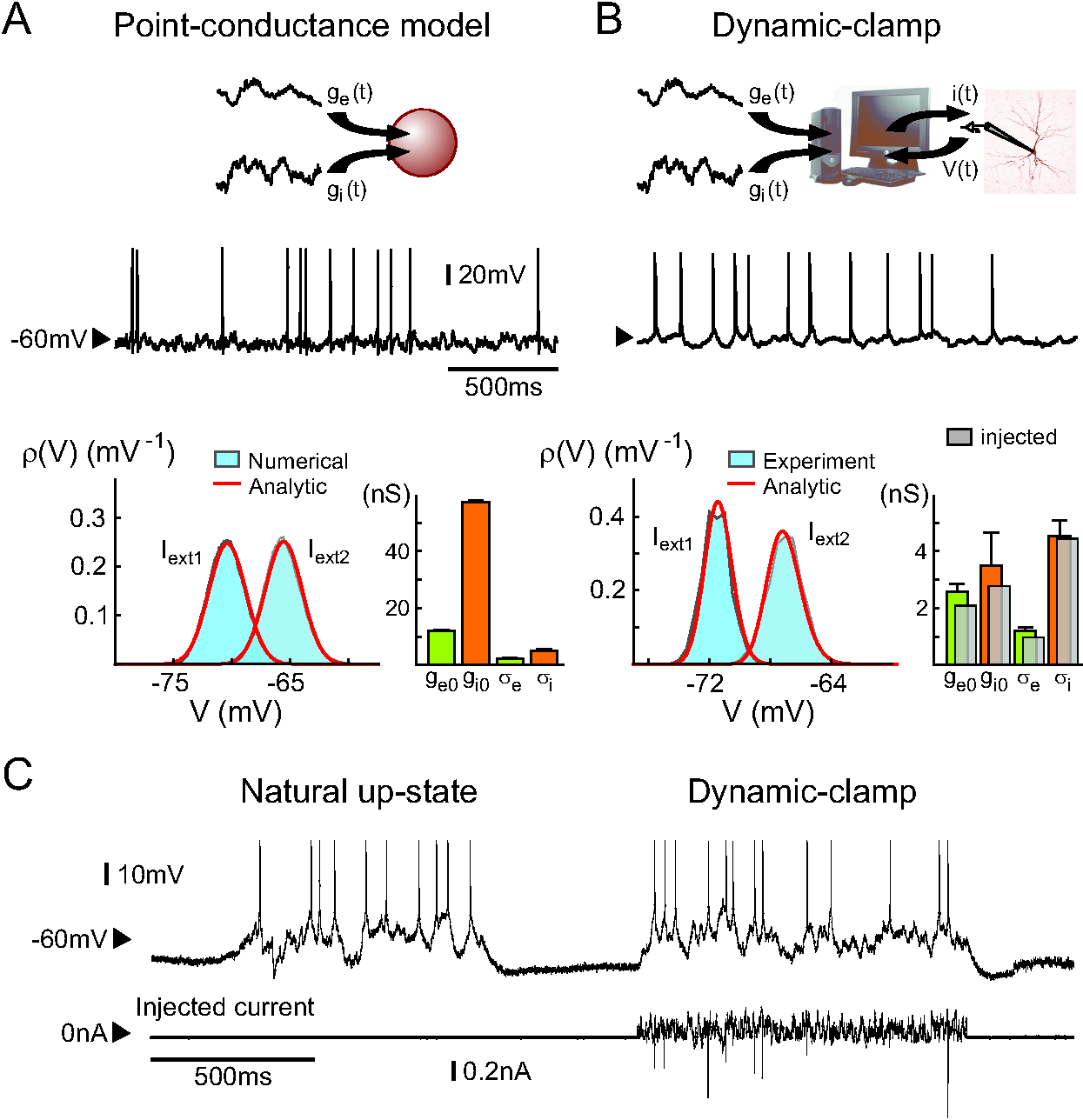}}

\caption{Conductance extraction from V$_m$ distributions: numerical
and dynamic-clamp test of the method. A. Simulation of the
point-conductance model (top trace) and comparison between
numerically computed V$_m$ distributions (bottom; blue) and the
analytic expression (red curves; conductance values shown in the bar
graph).  B.  Dynamic-clamp injection of the point-conductance model
in a real neuron. (Right) Conductance parameters are re-estimated
(back, colored; error bars are standard deviations obtained when the
same injected conductance parameters are re-estimated in different
cells) from the V$_m$ distributions and compared to the known
parameters of the injected conductances (front, grey). (Left) The
experimental V$_m$ distributions are compared to the analytic
distributions calculated using the re-estimated conductance
parameters. C.  Comparison of a spontaneous up-state (Natural
up-state) with an artificial up-state recreated using conductance
injection (Dynamic-clamp). Modified from Rudolph et al., 2004.}

\label{fig1}
\efi 

\bfi 
\centerline{\includegraphics[width=12cm]{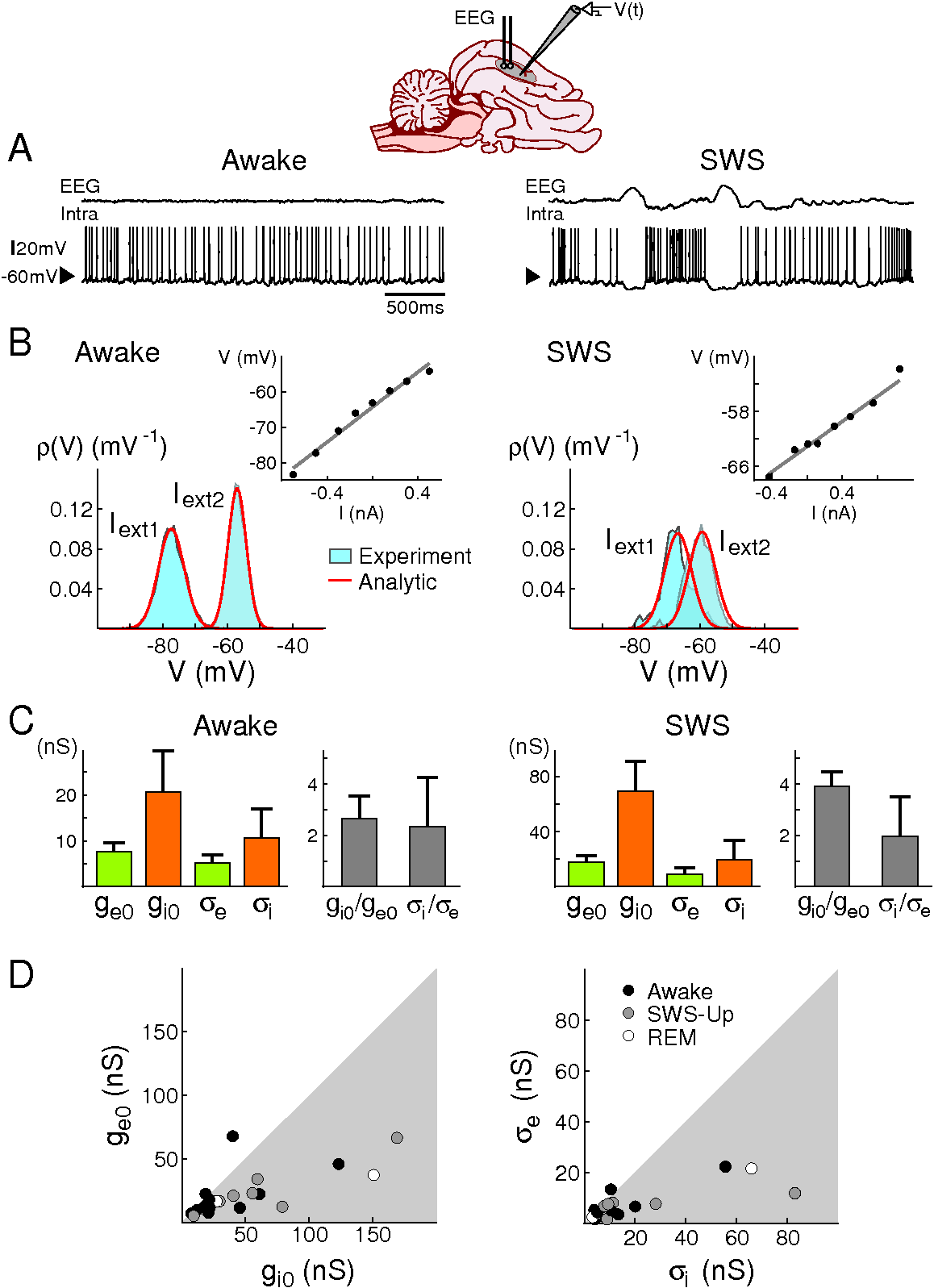}}

\caption{VmD estimation of conductances from intracellular recordings
in awake and naturally sleeping cats. A. Intracellular recordings in
awake and naturally sleeping (SWS) cats. Recordings were made in
association cortex (area 5-7). B. Examples of V$_m$ distributions
computed during wakefulness (Awake) and slow-wave sleep up-states
(SWS). The continuous lines show Gaussian fits of the experimental
distributions. Insets: current-voltage relations obtained for these
particular neurons. C. Conductance values estimated using the VmD
method. Results for the means ($g_{e0}$, $g_{i0}$) and standard
deviations ($\sigma_e$, $\sigma_i$) of excitatory and inhibitory
conductances, respectively, as well as their ratios are shown (error
bars: standard deviations obtained by repeating the analysis using
different pairs of injected current levels).  D. Grouped data showing
the means and standard deviations of the conductances for different
cells across different behavioral states (REM = Rapid Eye Movement
sleep). Figure modified from Rudolph et al., 2007.}

\label{fig2}
\efi 

\bfi 
\centerline{\includegraphics[width=11cm]{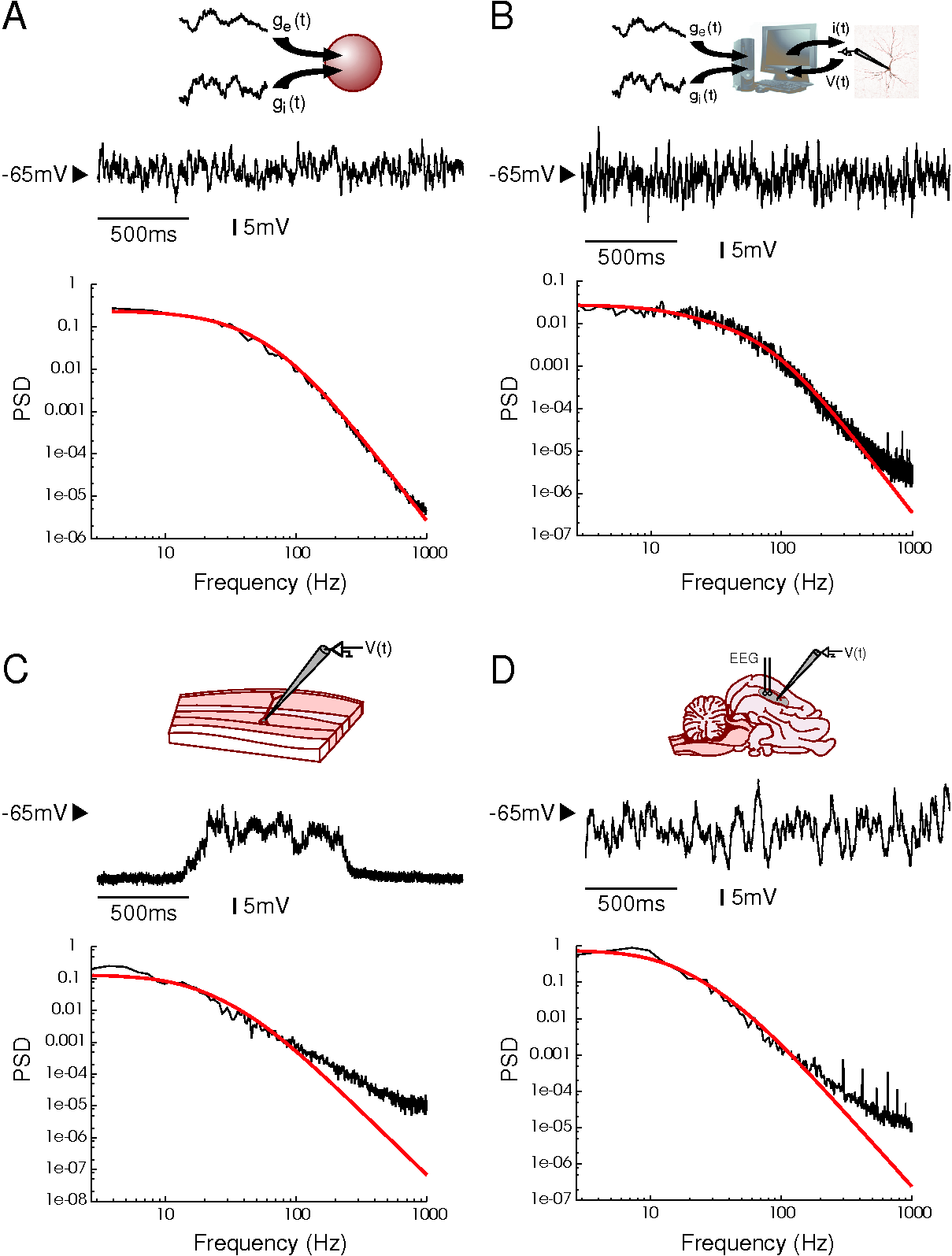}}

\caption{Fit of the synaptic time constants to the power spectrum of
the membrane potential. A. Comparison between the analytic prediction
(Eq.~\ref{VmPSD}; red) and the PSD of the V$_m$ for a
single-compartment model (Eq.~\ref{PCmodel}; black) subject to
excitatory and inhibitory fluctuating conductances
(Eqs.~\ref{flucte}-\ref{flucti}; $\tau_e$ = 3~ms and $\tau_i$ =
10~ms).  B. PSD of the V$_m$ activity in a guinea-pig visual cortex
neuron (black), where the same model of fluctuating conductances as
in A was injected using dynamic-clamp.  The red curve shows the
analytic prediction using the same parameters as the injected
conductances ($\tau_e$ = 2.7~ms and $\tau_i$ = 10.5~ms).  C. PSD of
V$_m$ activity obtained in a ferret visual cortex neuron (black)
during spontaneously occurring up-states. The PSD was computed by
averaging PSDs calculated for each up-state. The red curve shows the
best fit of the analytic expression with $\tau_e$ = 3~ms and $\tau_i$
= 10~ms.  D. PSD of V$_m$ activity recorded in cat association cortex
during activated states {\it in vivo}. The red curve shows the best
fit obtained with $\tau_e$ = 3~ms and $\tau_i$ = 10~ms. Panel A
modified from Destexhe and Rudolph, 2004; Panel D modified from
Rudolph et al., 2005.}

\label{fig3}
\efi 

\bfi 
\centerline{\includegraphics[width=14cm]{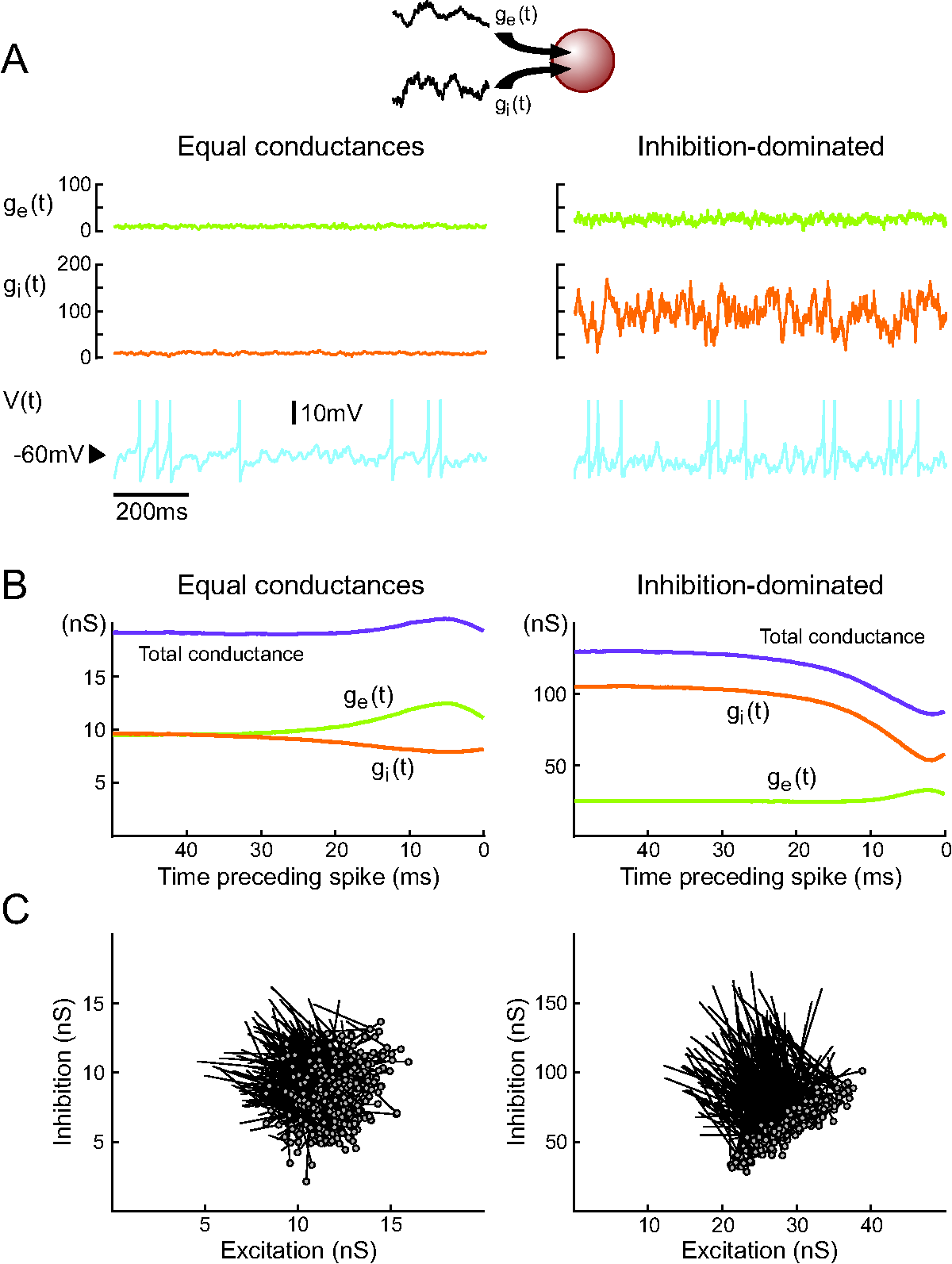}}

\caption{Comparison between equal conductances and
inhibition-dominated states in a computational model. A. Equal
conductance (left; $g_{e0}$ = $g_{i0}$ = 10~nS, $\sigma_e$ =
$\sigma_i$ = 2.5~nS) and inhibition-dominated states (right; $g_{e0}$
= 25~nS, $g_{i0}$ = 100~nS, $\sigma_e$ = 7~nS and $\sigma_i$ = 28~nS)
in the point-conductance model. Excitatory and inhibitory
conductances, and the membrane potential, are shown from top to
bottom. Action potentials (truncated here) were described by
Hodgkin-Huxley type models (Destexhe et al., 2001;
Eq.~\ref{PCmodel2}).  B.  Average conductance patterns triggering
spikes.  Spike-triggered averages (STAs) of excitatory, inhibitory
and total conductance were computed in a window of 50~ms before the
spike. C.  Vector representation showing the variation of synaptic
conductances preceding each spike.  The excitatory and inhibitory
conductances were averaged in two windows of 30-40~ms and 0-10~ms
(circle) before the spike, and a vector was drawn between the
obtained values.}

\label{fig4}
\efi 

\bfi 
\centerline{\includegraphics[width=14cm]{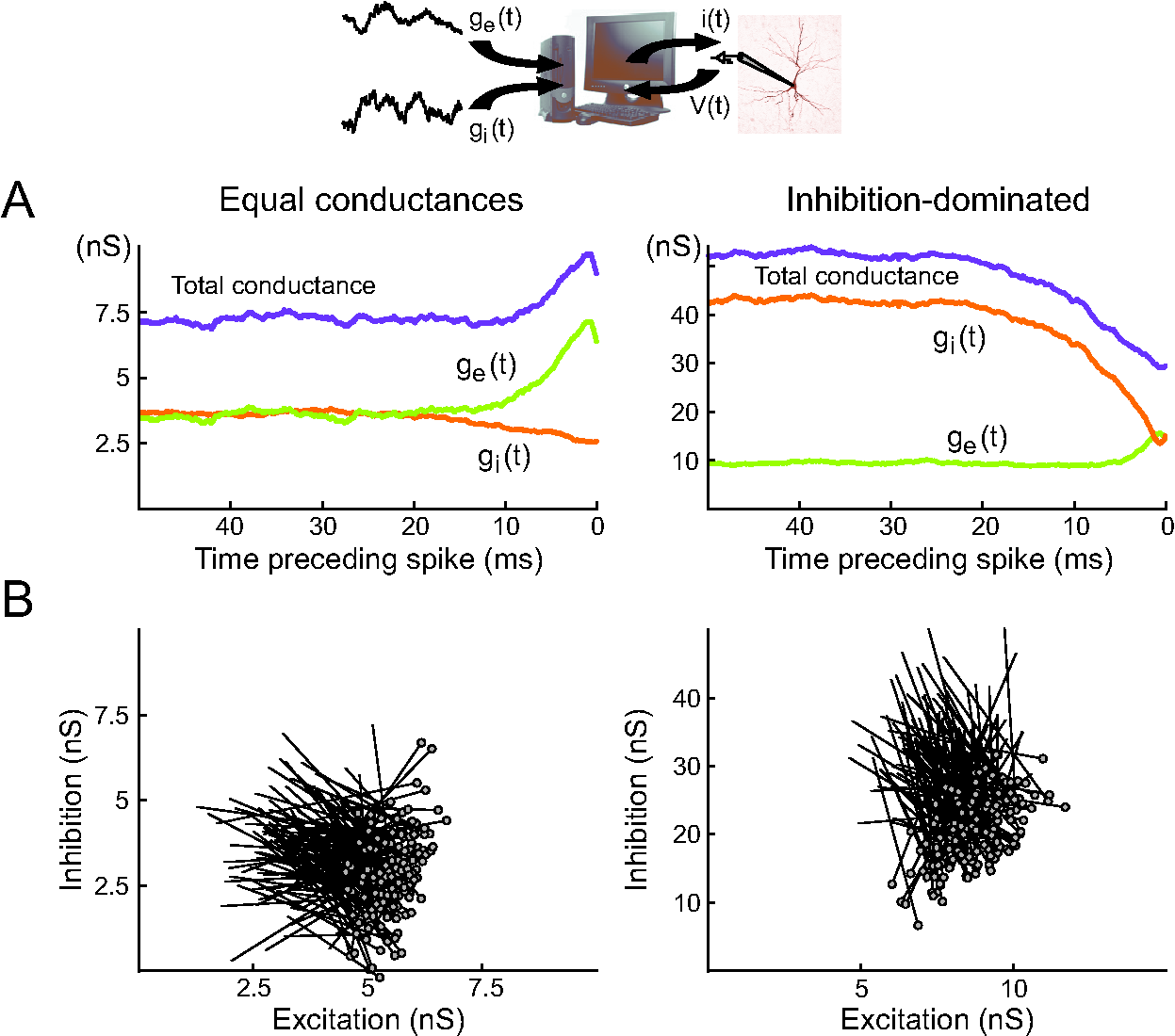}}

\caption{Average conductance patterns triggering spikes in
dynamic-clamp experiments. A. Spike-triggered averages of excitatory,
inhibitory and total conductance in a window of 50~ms before the
spike in a cortical neuron subject to fluctuating conductance
injection. The two states, equal conductances (left) and
inhibition-dominated (right), were recreated similar to the model
of Fig.~\ref{fig4}. Conductance STAs showed qualitatively
similar patterns. B. Vector representation showing the variation of
synaptic conductances preceding each spike (as in Fig.~\ref{fig4}C).}

\label{fig5}
\efi 

\bfi 
\centerline{\includegraphics[width=11cm]{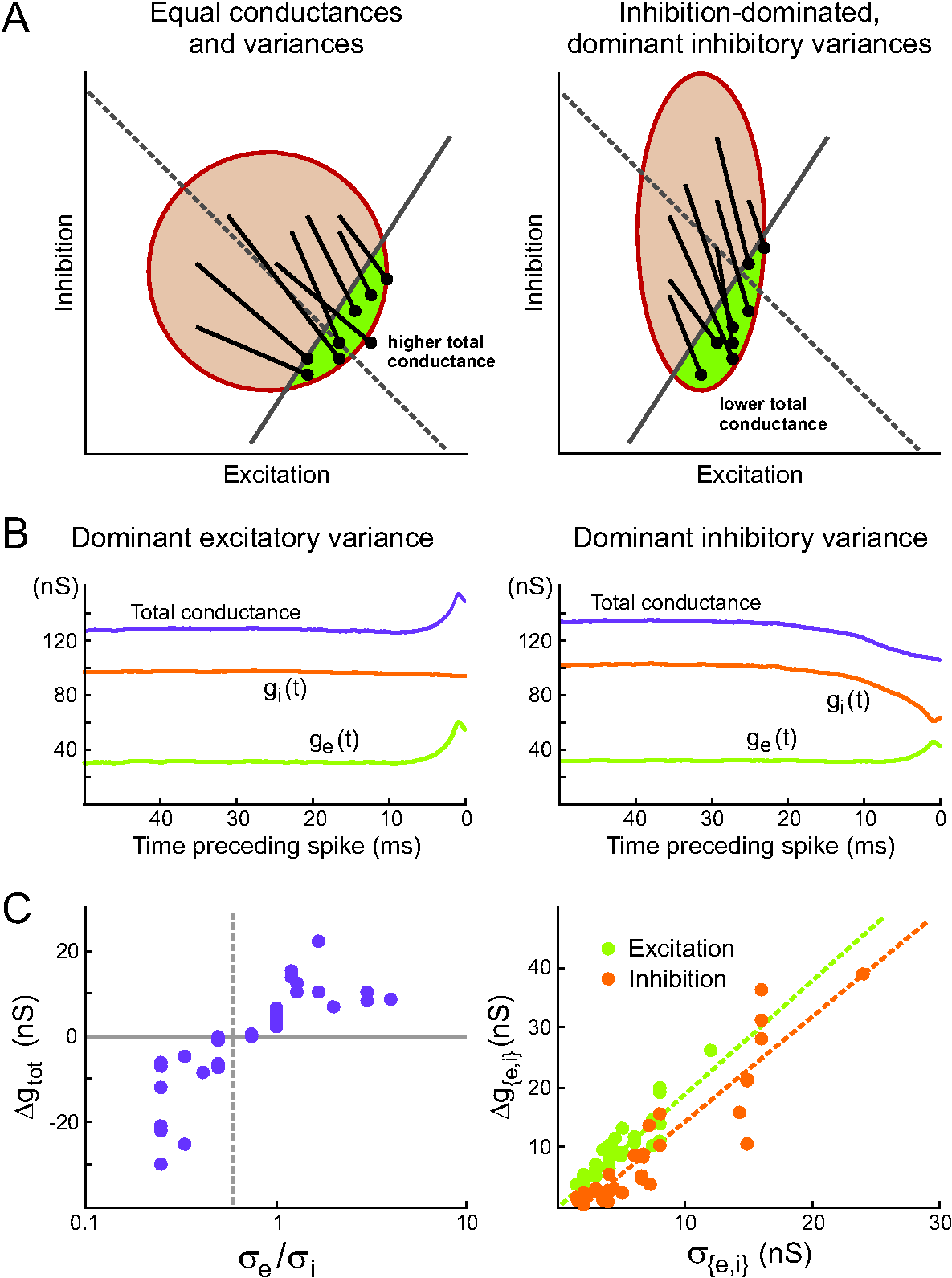}}

\caption{Geometrical interpretation of the average conductance
patterns preceding spikes and test in dynamic-clamp. A. Red ellipses:
isoprobable conductance configurations. Green area: conductance
configurations for which the total synaptic current is positive at
spike threshold. The vector representations of
Figs.~\ref{fig4}--\ref{fig5} are schematized here, and compared to
the lines defined by $\left\{g_e(E_e-V_t) + g_i(E_i-V_t) +
G_l(E_L-V_t)=0\right\}$ \corr{(solid gray line)} and
$\left\{g_e+g_i=g_{e0}+g_{i0}\right\}$ \corr{(dashed gray line)}. 
The angle between the two lines \corr{and the aspect ratio of the
ellipse} determine whether spikes are preceded, on average, by total
conductance increase (left) or decrease (right) (see text for further
explanations). B.  Spike-triggered average conductances obtained in
dynamic-clamp, illustrating that for the same average conductances,
the variances determine whether spikes are preceded by total
conductance increase (left) or decrease (right). C.  Geometrical
prediction tested in dynamic-clamp (left): grouped data showing total
conductance change preceding spikes as a function of the ratio
$\sigma_e$/$\sigma_i$.  The dashed line ($\sigma_e$/$\sigma_i$=0.6)
visualizes the predicted value separating total conductance increase
cases from total conductance decrease cases. In addition (right),
dynamic-clamp data indicates that the amplitude of change of each of
the conductances before a spike is linearly correlated with the
standard deviation parameter used for this conductance.}

\label{fig6}
\efi 

\bfi 
\centerline{\includegraphics[width=12cm]{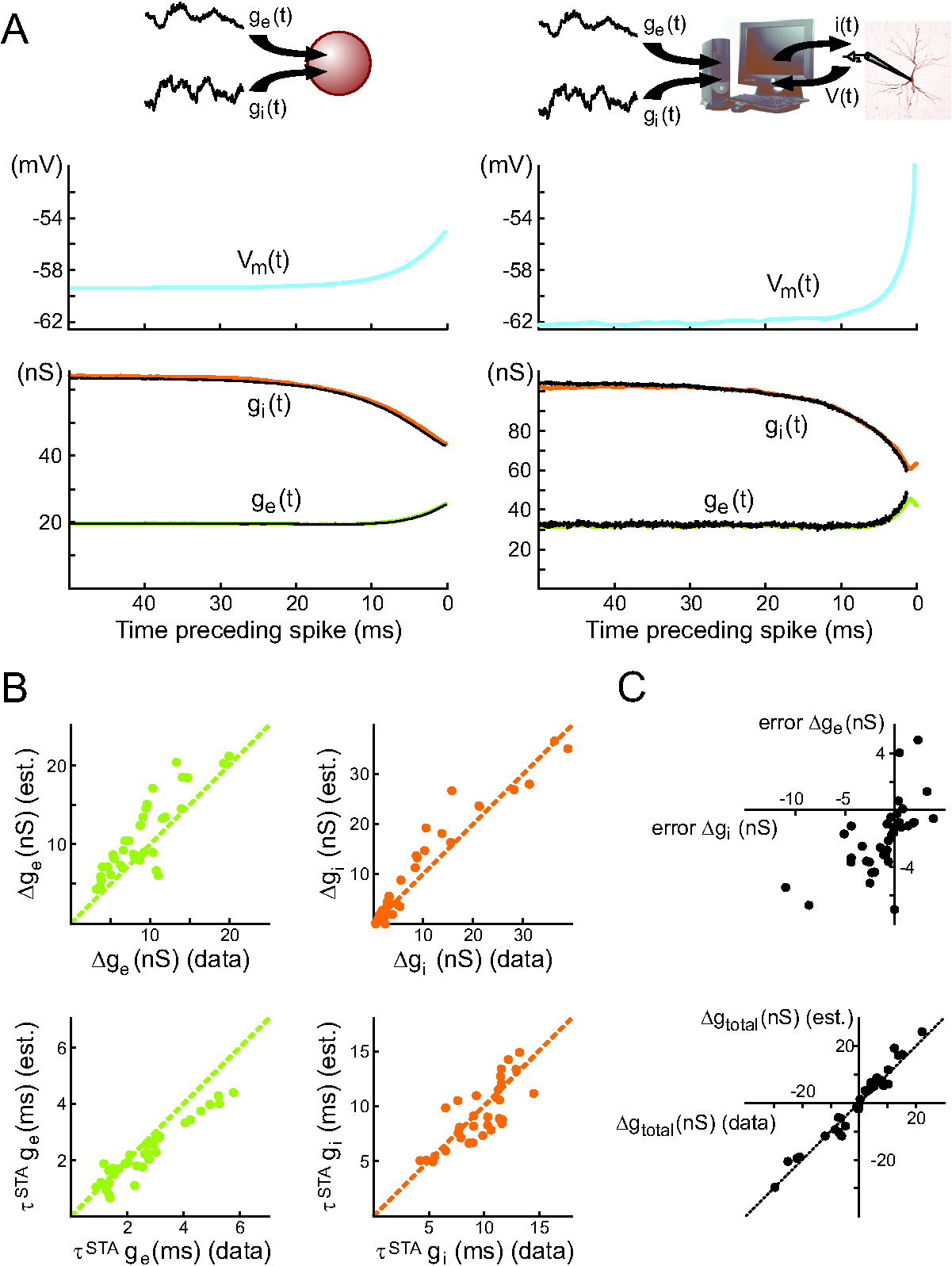}}

\caption{Spike-triggered conductance extraction from intracellular
recordings of the V$_m$ and test in dynamic-clamp. A. Left:
spike-triggered average (STA) of the V$_m$ in an integrate-and-fire
extension of the point-conductance model (top trace). The numerically
obtained conductance STAs (orange, green) are compared to the
conductance STAs extracted from the V$_m$ (black) (bottom trace). 
Right: test of the STA method using dynamic-clamp. The STA of the
V$_m$ is obtained following injection of fluctuating conductances
(top trace). The measured conductance STAs (orange, green) are
compared to the conductance STAs extracted from the V$_m$ (black)
(bottom trace). B. Grouped data comparing conductance STAs extracted
using the method with the conductance STAs measured following
dynamic-clamp injection: amplitude of conductance change preceding
the spike (top graphs) and time constant of this change (bottom
graphs), for both excitation and inhibition. C. Top: correlation
between errors for excitation and inhibition on the absolute value of
conductance variation. Bottom: total conductance change preceding
spikes; comparison between extracted and measured STAs. Dashed lines:
$Y=X$.}

\label{fig7}
\efi 

\bfi 
\centerline{\includegraphics[width=14cm]{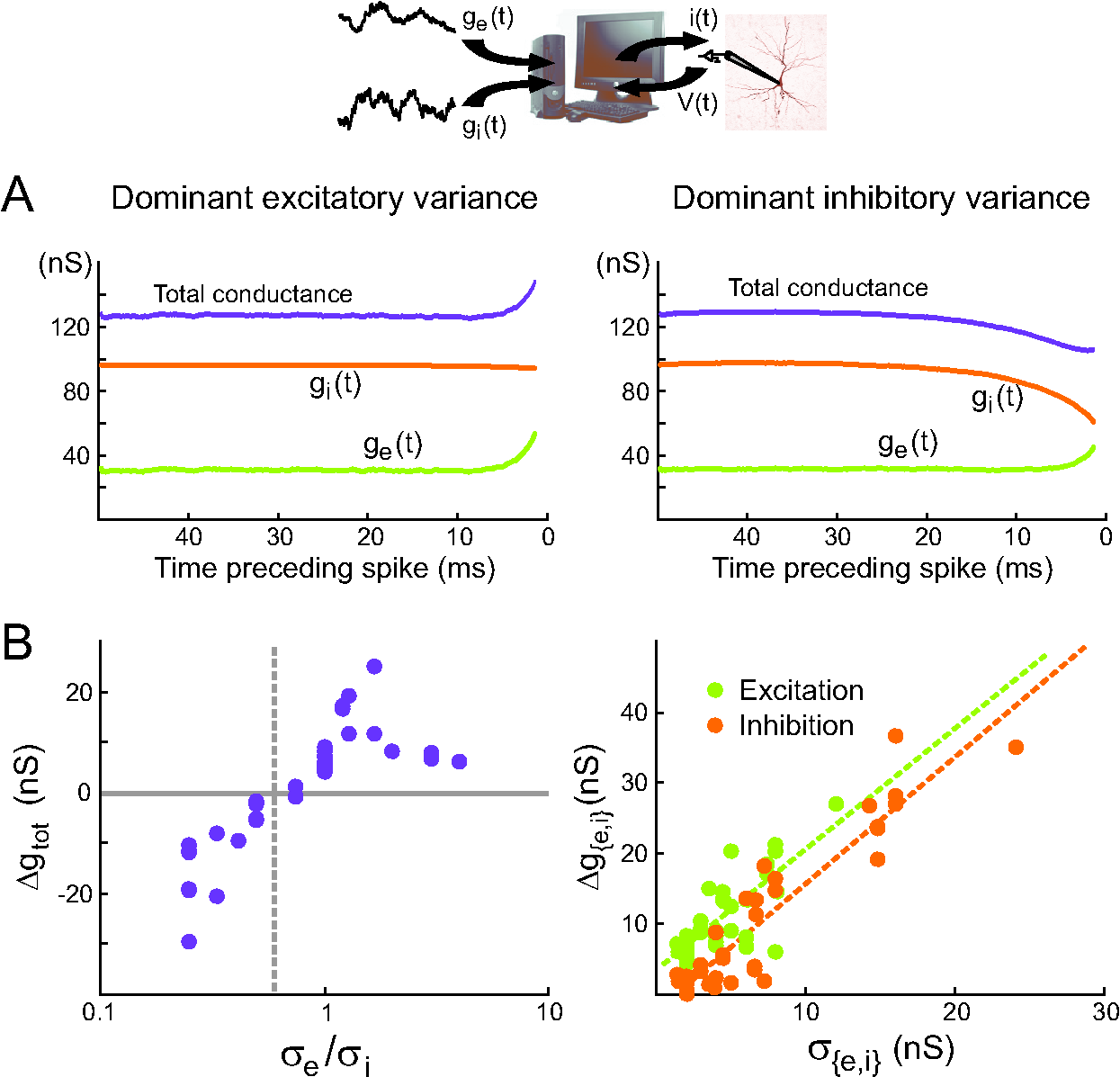}}

\caption{Analysis of the average conductance patterns preceding
spikes, same analysis as Fig.~\ref{fig6}B-C, but using conductance
STAs extracted from the V$_m$, instead of the measured ones. A. 
Example conductance STAs extracted from the V$_m$ STAs of the same
cell and the same conductance injections as Fig.~\ref{fig6}B. B. 
Left: test of the geometrical prediction (dashed line) using
conductance STAs extracted from the V$_m$, and showing total
conductance change preceding spikes as a function of the ratio
$\sigma_e$/$\sigma_i$ (as in Fig.~\ref{fig6}C, left).  Right:
correlation between the amplitude of change of each conductance
preceding a spike, as extracted from the V$_m$, and the standard
deviation parameter for this conductance (compare to
Fig.~\ref{fig6}C, right).}

\label{fig8}
\efi 

\bfi 
\centerline{\includegraphics[width=12cm]{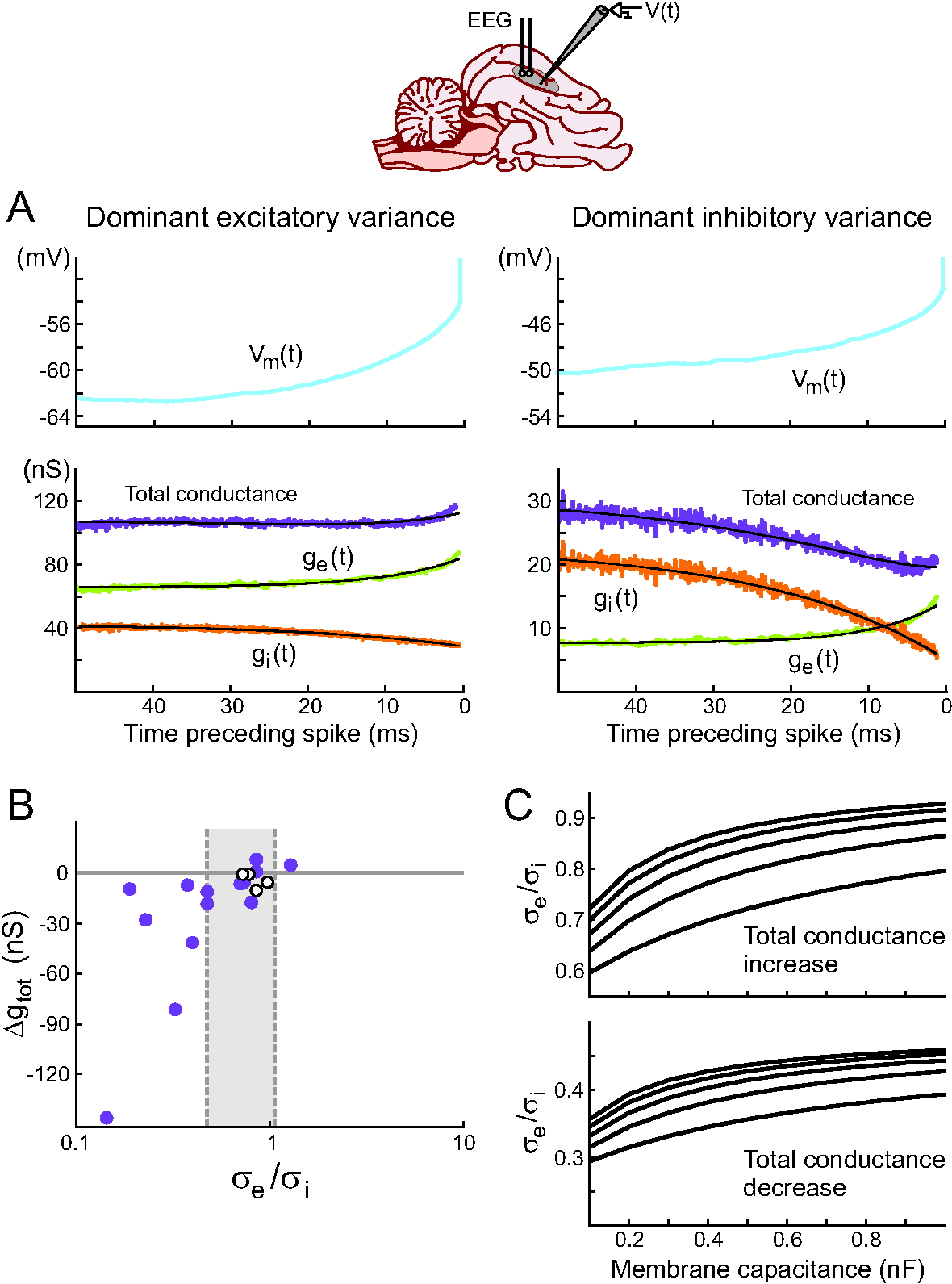}}

\caption{Spike-triggered conductance analysis {\it in vivo}.  A. STA
conductance analysis from intracellular recordings in awake and
sleeping cats. Two example cells are shown during wakefulness, and
for each, the V$_m$ STA (top) and the extracted conductance STAs
(bottom) are shown. In the first cell (left), the total conductance
increases before the spike. In the second example cell (right), the
total conductance decreases before the spike (black traces are
exponential fits to the extracted STAs). B. Total conductance change
preceding spikes as a function of the ratio $\sigma_e$/$\sigma_i$. 
Given the cell-to-cell variability of observed spike thresholds, each
cell has a different predicted ratio separating total conductance
increase cases from total conductance decrease cases. The two dashed
lines ($\sigma_e$/$\sigma_i$=0.48 and $\sigma_e$/$\sigma_i$=1.07)
visualize the two extreme predicted ratios. Cells in white are the
ones not conforming to the prediction.  C.  Dependency of the ratio
$\sigma_e$/$\sigma_i$ estimated by the VmD method on the value of the
membrane capacitance $C$.  Two sets of realistic V$_m$ distribution
parameters were used as input for the estimation, one leading to
$\sigma_e$/$\sigma_i$ $>$ 0.6 (left), another leading to
$\sigma_e$/$\sigma_i$ $<$ 0.6 (right). For each set, the total input
resistance was varied from 10~M$\Omega$ (bottom curves) to
50~M$\Omega$ (top curves), in steps of 10~M$\Omega$.  Panel A
modified from Rudolph et al., 2007.}

\label{fig9}
\efi 

\end{document}